\documentclass[preprintnumbers, floatfix, letterpaper, singlecolumn,aps,prd,epsfig,nofootinbib,natbib,
longbibliography
]{revtex4-2}
\usepackage{amsmath}
\newcommand\beq{\begin{eqnarray}}
\newcommand\eeq{\end{eqnarray}}

\def\lsim{\mathrel{\rlap{\lower4pt\hbox{$\sim$}}
    \raise1pt\hbox{$<$}}}                
\def\gsim{\mathrel{\rlap{\lower4pt\hbox{$\sim$}}
    \raise1pt\hbox{$>$}}}            

\allowdisplaybreaks
\usepackage{wrapfig}
\usepackage{graphicx}
\usepackage{setspace}
\usepackage{placeins}
\usepackage[T1]{fontenc} 
\usepackage{multirow}
\usepackage[table,xcdraw]{xcolor}
\usepackage[colorlinks,linkcolor=magenta,citecolor=magenta,urlcolor=magenta]{hyperref}
\usepackage{float}
\usepackage{subfigure}
\usepackage{placeins}
\usepackage{caption}
\usepackage{amsmath}
\usepackage[table,xcdraw]{xcolor}
\usepackage[utf8]{inputenc}
\usepackage{color}
\usepackage{amsfonts}
\usepackage{amssymb}
\usepackage{graphicx}
\begin{document}

\renewcommand{\theequation}{\arabic{section}.\arabic{equation}}
\renewcommand{\thefigure}{\arabic{section}.\arabic{figure}}
\renewcommand{\thetable}{\arabic{section}.\arabic{table}}

\newcommand{\orcid}[1]{\href{https://orcid.org/#1}{\includegraphics[width=10pt]{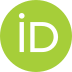}}}
\newcommand{\PRL}{Phys. Rev. Lett.}
\newcommand{\PLB}{Phys. Lett. B}
\newcommand{\PRD}{Phys. Rev. D}
\newcommand{\CQG}{Class. Quantum Grav.}
\newcommand{\JCAP}{J. Cosmol. Astropart. Phys.}
\newcommand{\JHEP}{J. High. Energy. Phys.}
\title{\Large \baselineskip=10pt 
  Inflationary Dynamics in Einstein-Gauss-Bonnet Gravity Using New Slow-Roll Approximations considering Generalised Reheating}
\author{Yogesh\orcid{0000-0002-7638-3082}}
\email{ yogesh@zjut.edu.cn, yogeshjjmi@gmail.com}
\affiliation{Institute for Theoretical Physics and Cosmology,
Zhejiang University of Technology, Hangzhou, 310023, China}

\author{Imtiyaz Ahmad Bhat\orcid{0000-0002-2695-9709}}
\email{imtiyaz@ctp-jamia.res.in}
\affiliation{Department of Physics, Islamic University of Science and Technology, Awantipora, J\&K- 192122 }

\author{Mayukh R. Gangopadhyay\orcid{0000-0002-1466-8525}}
\email{mayukh_ccsp@sgtuniversity.org, mayukhraj@gmail.com}
\affiliation{Centre for Cosmology and Science Popularization, SGT University, Gurugram, Haryana-122505, India}


\begin{abstract}
 The results of Cosmic Microwave Background (CMB) observations reported by {\it Planck} satellite suggest that a plateau characteristic in the flat potential of the inflaton field is favored to drive the Universe's early acceleration. On that note, Power Law Plateau (PLP) potential has gained a lot of attention. In this study, we demonstrate that implementing this inflationary model in an Einstein-Gauss-Bonnet(EGB) gravity background makes the model compatible with observations while avoiding late-stage thermal inflation, which is otherwise unavoidable in a standard scenario. More importantly, we have taken the PLP model as a case study to check the new slow-roll approximations method proposed in  \cite{Pozdeeva:2024ihc}. Then we take one step further to check the consequence of generalized reheating in this scenario. Thus opening a new promising window to study inflationary dynamics in EGB gravity.
\end{abstract}

\maketitle

\vspace{0.0001in}


\baselineskip=15.4pt

\section{Introduction\label{sec:intro}}
The inflationary epoch, along with the standard hot Big Bang model, could resolve the understanding of the birth and evolution of our Universe to its current state \cite{Liddle:2000cg}, \cite{1990eaun.book.....K}. The first idea of inflation was proposed by Guth \cite{Guth:1980zm}, and since then, the inflationary epoch has gained tremendous attention, becoming an interdisciplinary study requiring both the fields of high energy physics and cosmology\cite{Starobinsky:1980te,Linde:1981mu,Mukhanov:1981xt,Sato:1981qmu,1996tyli.conf..771S,PhysRevLett.48.1220,Starobinsky:1982ee}. CMB observations by the {\it Planck} and Wilkinson Microwave Anisotropy Probe(WMAP) missions have ruled out or strongly disfavored most of the textbook models of inflation in the standard scenario\cite{Hinshaw_2013, Planck:2015sxf,Planck:2018jri}. Based on the current observation, it becomes evident that a plateau-like feature in the potential can satisfy the constraints imposed by CMB in the best possible manner. In a key study, Dimopoulos and Owen \cite{Dimopoulos:2016zhy} presented the Power Law Plateau inflation (PLP) with plateau properties. However, in this theoretically justified model, one needs to incorporate a phase of thermal inflation to allow it to withstand the observational constraints while remaining in the conventional cold inflationary premise.

 It was shown for the first time in \cite{Adhikari:2020xcg} that if one wants to develop the inflationary models under the UV complete string theory, PLP in the alternative theory of Randall-Sundrum brane-world\cite{Randall:1999ee, Randall:1999vf}can be compatible with the observation, and  it can circumvent the tighter constraints from string theory, such as swampland conjecture\cite{Vafa:2005ui}. However, on the other hand, RS models suffer serious existential crises because of the continuous limitation on the scale provided by the braneworld theory\cite{Bhattacharya:2019ryo}.  It has been shown that in modified gravity, the restriction on the inflationary models can be less stringent\cite{Kallosh:2013pby,1982PhLB..117..175S,Starobinsky:1983zz,Barvinsky:1994hx,Cervantes-Cota:1995ehs,Bezrukov:2007ep,Barvinsky:2008ia,DeSimone:2008ei,Bezrukov:2008ej,Barvinsky:2009ii,Bezrukov:2010jz,Bezrukov:2013fka,Rubio:2018ogq,Koshelev:2020xby,Elizalde:2014xva,Canko:2019mud,Gialamas:2020snr,Gialamas:2020vto,Gialamas:2021enw}. In other words, models that are disfavoured by observations in the standard cold inflationary model can satisfy the observational constraints in the modified gravity. The addition of the Gauss-Bonnet component to the Einstein-Hilbert action does not affect the equations of motion, as it is a total derivative. When EGB coupling $\xi(\phi)$ is combined with a function of the scalar field $\phi$, it becomes dynamically important. In string theory, the Gauss-Bonnet term serves as a quantum correction to the Einstein-Hilbert action. Numerous inflationary models have been studied in this context\cite{vandeBruck:2015gjd,Guo:2009uk,Guo:2010jr,Koh:2016abf,Pozdeeva:2020shl,Satoh:2008ck,Jiang:2013gza,Koh:2014bka,Koh:2018qcy,Mathew:2016anx,Mathew:2016anx,Pozdeeva:2020apf,Pozdeeva:2016cja,Nozari:2017rta,Armaleo:2017lgr,Yi:2018gse,Yi:2018dhl,Odintsov:2018zhw,Fomin:2019yls,Fomin:2020hfh,Kleidis:2019ywv,Rashidi:2020wwg,Odintsov:2020sqy,Odintsov:2020zkl,Kawai:2021bye,Kawai:2017kqt,Oikonomou:2022xoq,Oikonomou:2022ksx,Cognola:2006sp,Odintsov:2020xji,Odintsov:2020mkz,Oikonomou:2020sij,Nojiri:2019dwl,Fomin:2019yls,Ashrafzadeh:2024oll,Solbi:2024zhl,Ashrafzadeh:2023ndt,Oikonomou:2024etl,Oikonomou:2024jqv,Odintsov:2023weg,Kawai:2023nqs,Kawai:1999pw,Kawai:1998ab,Yogesh:2024zwi,Yogesh:2025hll,Pozdeeva:2024kzb}; the most widely studied EGB coupling is the inverse function of the scalar field\cite{Guo:2009uk,Guo:2010jr,Koh:2016abf,Pozdeeva:2020shl,Jiang:2013gza,Yi:2018gse,Odintsov:2018zhw,Kleidis:2019ywv,Rashidi:2020wwg}. However, in this manuscript, we will study a different form of coupling that involves a $``\tanh"$ term\cite{Kawai:2021edk}. In \cite{Pozdeeva:2024ihc}, the authors have presented a new kind of slow-roll approximations, which are found to be more accurate in comparison to the previous slow-roll approximations presented in\cite{Pozdeeva:2020apf}. As it has been already shown in \cite{Pozdeeva:2024ihc} that new slow approximations are more accurate, we aim to do a case study of the new slow-roll approximations in the EGB scenario using the the PLP potential as a prototype. This serves the primary objective of this exercise. More importantly, we extend the study to accommodate the generalized reheating and review its effect in the effective EGB background. The standard cold inflationary scenario requires an epoch of reheating after the end of inflation to enter the radiation-dominated epoch and start the Big Bang Nucleosynthesis (BBN). Studying the reheating phase following the end of inflation is necessary for the conceptual development of inflationary model construction. We have studied reheating in this work and established the limits on the number of e-folding and reheating temperatures that are permitted by CMB data.

 The paper is organized as follows: In section~\ref{EGB}, we will go over some
of the fundamental critical features in the EGB framework. A theoretically well-motivated Power Law Plateau (PLP) potential for this study has been reviewed in section~\ref{plp}. In section~\ref{sraw}, the calculation of slow-roll parameters without any approximations is reviewed. In section~\ref{sranew}, the new slow-roll approximations (SRA) as reported in the work~\cite{Pozdeeva:2024ihc} are reviewed. Finally, the analysis of the inflationary dynamics using the new SRA and the effect of generalized reheating has been reported in the sections \ref{infa} and \ref{sec:reh}. Finally, we conclude in section~\ref{discussions} with our conclusions and future direction of work.
\section{Review of Einstein-Gauss Bonnet Gravity}
\label{EGB}
 
The effective action of the Gauss-Bonnet term is described by the following action:
\begin{equation}
\label{action1}
S=\int d^4x\sqrt{-g}\left[U_0R-\frac{1}{2}g^{\mu\nu}\partial_\mu\phi\partial_\nu\phi-V(\phi)-\frac{1}{2}\xi(\phi){\cal G}\right],
\end{equation}
where $U_0>0$ is a constant, $V(\phi)$ and $\xi(\phi)$ are the potential and EGB coupling and differentiable functions. $R$ denotes the usual Ricci scalar. The EGB term is given by: 
\begin{equation}
   \mathcal{G}=R_{\mu\nu\rho\sigma}R^{\mu\nu\rho\sigma}-4R_{\mu\nu}R^{\mu\nu}+R^2. 
\end{equation}

The evolution of equations and dynamical systems in Einstein-Gauss-Bonnet gravity can be written for a spatially flat FLRW metric
\cite{vandeBruck:2015gjd,Pozdeeva:2020apf}as:
\begin{eqnarray}
12H^2\left(U_0-2\xi_{,\phi}\psi H\right)&=&\psi^2+2V, 
\label{Equ00} \\
4\dot H\left(U_0-2\xi_{,\phi}\psi H\right)&=&{}-\psi^2+4H^2\left(\xi_{,\phi\phi}\psi^2+\xi_{,\phi}\dot\psi-H\xi_{,\phi}\psi\right), 
\label{Equ11} \\
\dot{\psi}+3H\psi&=&{} -V_{,\phi} -12H^2\xi_{,\phi}\left(\dot{H}+H^2\right),\label{Equphi}
\end{eqnarray}
where $H=\dot{a}/a$ denotes the Hubble parameter and  $a(t)$ is the scale factor, the dot stands for the time derivative here. The term $\psi=\dot{\phi}$ and $A_{,\phi} \equiv \frac{dA}{d\phi}$ (true for any function $A(\phi)$).

    Using the relation ${d}/{dt}=H\, {d}/{dN}$ and introducing $\chi=\psi/H$, one can write system. (For detailed calculation see~\cite{Pozdeeva:2024ihc}):
\begin{equation}
\label{DynSYSN}
\begin{split}
\frac{d\phi}{dN}=&\,\chi,\\
\frac{d\chi}{dN}=&\,\frac{1}{Q\left(B-2\xi_{,\phi}Q\chi\right)}\left\{
3\left[3-4\xi_{,\phi\phi} Q\right]\xi_{,\phi}Q^2\chi^2+ \left[3B+2\xi_{,\phi}V_{,\phi}-6U_0\right]Q\chi-\frac{V^2}{U_0}X\right\}\\
&{}-\frac{\chi}{2Q}\frac{dQ}{dN},\\
\frac{dQ}{dN}=&\,\frac{Q}{2\left(B-2\xi_{,\phi}Q\chi\right)}\left\{\left(4\xi_{,\phi\phi}Q-1\right)\chi^2-16\xi_{,\phi}Q\chi-4\frac{V^2}{U_0^2}\xi_{,\phi}X\right\},
\end{split}
\end{equation}

where $Q\equiv H^2$, $B=12\xi_{,\phi}^2H^4+U_0$ and $X=\frac{U_0^2}{V^2}\left(12\xi_{,\phi} H^4+V_{,\phi} \right)$

    Equation (\ref{Equ00}) can be presented in the following form
\begin{equation}
\label{Equ00N}
24\xi_{,\phi}\chi Q^2+\left(\chi^2- 12U_0\right)Q+2V=0.
\end{equation}
and has solutions:
\begin{equation}
\label{Q0}
Q_\pm=\frac{12U_0-\chi^2\pm\sqrt{\left(12U_0-\chi^2\right)^2-192V\xi_{,\phi}\chi}}{48\xi_{,\phi}\chi},
\end{equation}
if $\xi_{,\phi}\chi\neq 0$. In the opposite case, the unique solution exists
\begin{equation}
Q_0=\frac{2V}{12U_0-\chi^2}\,.
\end{equation}
Equation (\ref{Equ00N}) restricts the set of initial conditions of system~(\ref{DynSYSN}).

\section{Reviewing PLP inflation}
\label{plp}
    
    As the most advanced extension of space-time symmetries permitted by the Coleman–Mandula no-go theorem \cite{Coleman:1967ad}, supersymmetry goes beyond Poincaré symmetry. Within this setting, and as shown in \cite{Dimopoulos:2016zhy}, one can write the superpotential as;

\begin{equation}
W= \frac{S^2(\Phi^2_1- \Phi^2_2)}{2 m}~,
\label{eq1}
\end{equation}
Here, $S, \Phi_1, \Phi_2$ are the chiral superfields, while $m$ represents the corresponding sub-Planckian energy scale. From \cite{Dimopoulos:2016zhy}, the F-term scalar potential takes the form:
\begin{equation}
    V_F= \frac{|S|^2}{m^2} \Big[|\Phi^2_1-\Phi^2_2|^2+|S|^2 \left(|\Phi_1|^2 + |\Phi_2|^2   \right)  \Big]
    \label{VF1}~.
\end{equation}
The potential \ref{VF1} gets minimised for $\Phi_1= \Phi_2$ . After imposing an appropriate  R-symmetry and performing the field rotation in the configuration space, we can introduce the canonically normalised real scalar field $\phi$ such as $|\Phi_1|= |\Phi_2|\equiv \frac{1}{2} \phi$. This leads to the following;
\begin{equation}
    V_F = \frac{|S|^4 \phi^2}{2 m^2}.
    \label{VF2}
\end{equation}
After considering the D-term contribution to the scalar potential~\cite{delaMacorra:1995qh}, we are left with;
\begin{equation} 
V_D = \frac{1}{2} \left( |S|^2- \sqrt{2} M^2 \right)^2\;,
 \end{equation}
where $M$ signifies the scale of grad unified theory ($\text{GUT}$), the total potential can be denoted as:
\begin{equation}
    V = \frac{|S|^4 \phi^2}{2 m^2}+ \frac{1}{2} \left( |S|^2-\sqrt{2}M^2 \right)^2~.
    \label{vtotal}
\end{equation}
After imposing the condition that the potential is minimized with respect to the S direction
\begin{equation}
    \frac{\partial V}{\partial |S|} = 0 \;\Rightarrow\; \langle |S|^2\rangle = \frac{\sqrt{2}M^2}{1+\phi^2/m^2}~.
    \label{sdire}
\end{equation}
Substituting Eq.~\ref{sdire} into Eq. \ref{vtotal} we get:
\begin{equation}
    V = \frac{M^4 \phi^2}{m^2+\phi^2}~,
\end{equation}
and writing $M^4= V_0 M^4_P$, we get the following form of the potential
\begin{equation}
V{(\phi)}= V_0 M_P^4 \frac{\phi^2}{m^2 + \phi^2}~.
\label{eqq2}
\end{equation}

Following \cite{Dimopoulos:2016zhy} we obtained a more generalized form of the potential: 
\begin{equation}
V{(\phi)}= V_0 M_P^4 \left( \frac{\phi^n}{\phi^n + m^n}\right)^q~.
\label{eqq2}
\end{equation}
where $M_P$ signifies the Planck mass and the $\phi$ is the inflation field. $V_0$ is the dimensionless parameter that needs to be fixed by matching the amplitude of the scalar power spectrum at the pivot scale to $2.0989\times10^{-9}$. Two potential parameters $n$ and $q$ can vary from $1$ to $4$, leading to $16$ different cases. Eq.~(\ref{eqq2}) is the inflationary potential, which we are interested in for our study. It is obvious to see that, if $m$ is super-Planckian, then this leads to just a large field $\phi^m$ model class. Thus, keeping $m \leq 1$ in the Planck unit is necessary, for simplicity in our calculation, we take $m=1$. However, the analyses we present in this manuscript can easily be extended to other smaller values of $m$. The standard case has been studied in \cite{Dimopoulos:2016zhy}, and it required a phase of thermal inflation to make it viable from the observational point of view. Studying such models in modified gravity can alleviate the necessity of thermal inflation. 


\section{The slow-roll parameters without any approximation}
\label{sraw}
~~~~Following Refs.~\cite{Guo:2010jr,vandeBruck:2015gjd,Pozdeeva:2020apf,Odintsov:2023lbb}, we consider the slow-roll parameters:
\begin{equation}
\label{epsilon}
\varepsilon_1 ={}-\frac{\dot{H}}{H^2}={}-\frac{d\ln(H)}{dN},\qquad \varepsilon_{i+1}= \frac{d\ln|\varepsilon_i|}{dN},\quad i\geqslant 1,
\end{equation}
\begin{equation}
\label{delta}
\delta_1= \frac{2}{U_0}\xi_{,\phi}H\psi=\frac{2}{U_0}\xi_{,\phi}H^2\chi,\qquad \delta_{i+1}=\frac{d\ln|\delta_i|}{dN}, \quad i\geqslant 1.
\end{equation}
It is easy to see that
\begin{equation}
\label{delta2}
\delta_2=\frac{\dot{\psi}}{H\psi}+\frac{\xi_{,\phi\phi}\psi}{H\xi_{,\phi}}-\varepsilon_1.
\end{equation}

    Using system~(\ref{DynSYSN}), we obtain that the parameter $\varepsilon_1(N)$ satisfies the following equation:
\begin{equation}
\label{equepsilon1a}
\varepsilon_1={}-\frac{1}{2Q}\frac{dQ}{dN}={}-\frac{1}{4\left(B-2\xi_{,\phi}Q\chi\right)}\left\{\left(4\xi_{,\phi\phi}Q-1\right)\chi^2-16\xi_{,\phi}Q\chi-4\frac{V^2}{U_0^2}\xi_{,\phi}X\right\}\,.
\end{equation}
    The spectral index $n_s$ and the tensor-to-scalar ratio $r$ are connected with the slow-roll parameters as follows~\cite{Guo:2010jr},
\begin{equation}
\label{ns_slr}
n_s=1-2\varepsilon_1-\frac{2\varepsilon_1\varepsilon_2-\delta_1\delta_2}{2\varepsilon_1-\delta_1}=1-2\varepsilon_1-\frac{d\ln(r)}{dN}=1+\frac{d}{dN}\ln\left(\frac{Q}{U_0r}\right),
\end{equation}
\begin{equation}
\label{r_slr}
r=8|2\varepsilon_1-\delta_1|.
\end{equation}
After this, we can find the amplitude of the scalar power spectrum as:
\begin{equation}
\label{As_slr}
A_s=\frac{Q}{\pi^2 U_0\, r}\,.
\end{equation}

\section{New slow-roll approximations}
\label{sranew}
To examine the stability of de Sitter solutions in model (\ref{action1}), an effective potential approach has been proposed in ~\cite{Pozdeeva:2019agu} (see also,~\cite{Pozdeeva:2020apf,Vernov:2021hxo}):
\begin{equation}
\label{Veff}
V_{eff}(\phi)={}-\frac{U_0^2}{V(\phi)}+\frac{1}{3}\xi(\phi).
\end{equation}

\label{New-slow-roll-approximation}
\subsection{New approximation I}
We construct slow-roll approximations with. (For detailed calculation see~\cite{Pozdeeva:2024ihc})
\begin{equation}
\label{equ00slr}
Q\simeq\frac{V}{6\,U_0}\left(1+\delta_1\right)=\frac{1}{6U_0^2}\left[U_0V+2V\xi_{,\phi}H\psi\right]\,.
\end{equation}

Following ~\cite{Pozdeeva:2024ihc} we write the slow roll parameters as: 
\begin{equation}
\delta_1(\phi)={}-\frac{2\, V^2\xi_{,\phi}\,{V_{eff}}_{,\phi}}{V^2\xi_{,\phi}^2+3\,U_0^3}\,.
\label{delta1phi}
\end{equation}

\begin{equation}
\label{apprIeps1phi}
\varepsilon_1(\phi)={}-\frac{1}{2}\frac{d\phi}{dN}\frac{d\ln(Q)}{d\phi}=\frac{3U_0^2\left(3U_0^2V_{,\phi}+\xi_{,\phi}V^2\right)}{V\left(9U_0^3-6U_0^2\xi_{,\phi}V_{,\phi}+\xi_{,\phi}^2V^2\right)}\,\frac{d\ln(Q)}{d\phi}\\
\end{equation}
where
\begin{equation*}
\frac{d\ln(Q)}{d\phi}=\frac{V_{,\phi}}{V}+\frac{
2\left[\xi_{,\phi}\xi_{,\phi\phi}V^2+\xi_{,\phi}^2VV_{,\phi}-3U_0^2\xi_{,\phi\phi}V_{,\phi}-3U_0^2\xi_{,\phi}V_{,\phi\phi}\right]}
{9U_0^3-6U_0^2\xi_{,\phi}V_{,\phi}+\xi_{,\phi}^2V^2}-\frac{2\xi_{,\phi}\xi_{,\phi\phi}V^2+2{\xi{,\phi}}^2VV_{,\phi}}{3U_0^3+\xi_{,\phi}^2V^2}.
\end{equation*}

    Also, we get
\begin{equation}
\label{eps2delta2phi}
\varepsilon_2(\phi)=\frac{U_0\delta_1}{2\xi_{,\phi}Q\varepsilon_1}{\varepsilon_1}_{,\phi}\,,\qquad
\delta_2=\frac{U_0}{2Q\xi_{,\phi}}{\delta_1}_{,\phi}.
\end{equation}
The relation between the number of e-folds andthe  field can be written as~\cite{Pozdeeva:2024ihc}:
\begin{equation}
\label{apprIdNDdphi}
\frac{dN}{d\phi}= {}-\frac{V^2\xi_{,\phi}^2+3U_0^3-2V^2\xi_{,\phi}{V_{eff}}_{,\phi}}{6U_0^2V{V_{eff}}_{,\phi}}\,.
\end{equation}

\subsection{New approximation II}
The second way to get $Q$ following~\cite{Pozdeeva:2024ihc}
\begin{equation}
\label{apprIIH2}
Q=\frac{V}{6U_0(1-\delta_1)}.
\end{equation}

Slow roll parameters in approximation-II can be written as: 
\begin{equation}
\label{apprIIdel1phi}
\delta_1(\phi)={}-\frac{2\xi_{,\phi}\left(3U_0^2V_{,\phi}+V^2\xi_{,\phi}\right)}{9U_0^2\left(U_0-\xi_{,\phi}V_{,\phi}\right)}.
\end{equation}

\begin{equation}
\label{apprIIeps1phi}
\varepsilon_1(\phi)={}-\frac{1}{2}\frac{d\phi}{dN}\frac{d\ln(Q)}{d\phi}=\frac{\left(3U_0^2V_{,\phi}+\xi_{,\phi}V^2\right)\left(9U_0^3-3U_0^2\xi_{,\phi}V_{,\phi}+2\xi_{,\phi}^2V^2\right)}
{27U_0^2V{\left(U_0-\xi_{,\phi}V_{,\phi}\right)}^2}\,\frac{d\ln(Q)}{d\phi}\,,
\end{equation}
where 
\begin{equation}
\label{apprIIdH2dphi}
\frac{d\ln(Q)}{d\phi}=\frac{V_{,\phi}}{V}+\frac{\xi_{,\phi\phi}V_{,\phi}+\xi_{,\phi}V_{,\phi\phi}}{\xi_{,\phi}V_{,\phi}-U_0}+\frac{3U_0^2\xi_{,\phi\phi}V_{,\phi}+3U_0^2\xi_{,\phi}V_{,\phi\phi}-4\xi_{,\phi}\xi_{,\phi\phi}V^2
-4\xi_{,\phi}^2VV_{,\phi}}{9U_0^3-3U_0^2\xi_{,\phi}V_{,\phi}+2\xi_{,\phi}^2V^2}\,.
\end{equation}
Using \ref{eps2delta2phi}, one can obtain the remaining slow roll parameters($\varepsilon_2, \delta_2$). 
The relation between the number of e-folds and the field can be written as~\cite{Pozdeeva:2024ihc}:
\begin{equation}
\label{apprIIequdNdphi}
\frac{dN}{d\phi}={}-\frac{27U_0^2V{\left(U_0-\xi_{,\phi}V_{,\phi}\right)}^2}{2\left(3U_0^2V_{,\phi}+\xi_{,\phi}V^2\right)\left(9U_0^3-3U_0^2\xi_{,\phi}V_{,\phi}+2\xi_{,\phi}^2V^2\right)},
\end{equation}


\section{Inflationary analysis}
\label{infa}
In the subsequent sections, we will provide the calculation of inflationary observables for PLP in the new slow roll approximations. For our analysis, we choose the following form of the EGB coupling \cite{Khan:2022odn,Gangopadhyay:2022vgh,Yogesh:2025wak}
\begin{equation}
    \xi(\phi) = \frac{\xi_1}{V_0} \tanh \left( \xi_2 (\phi) \right)
    \label{xicoupling}
\end{equation}
where $\xi_1$, $\xi_2$ are the coupling constants and $V_0$ is the scale of inflation, which we fix from the scalar power spectrum. Substituting Eq.~(\ref{eqq2}) and Eq.~(\ref{xicoupling}) in Eq.~(\ref{Veff}), we can derive the explicit form of the effective potential:
\begin{equation}
    V_{eff}(\phi)= \frac{-U_0^2 \left( \frac{\phi^n}{\phi^n+ m^n} \right)^{-q}}{V_0} + \frac{\xi_1 \tanh \left( \xi_2 \phi \right)}{3 V_0}
    \label{Veff1}
\end{equation}
In our analysis, we take $U_0=\frac{1}{2}$ and $M_P=1$. 
\subsection{Inflationary Observables for $\xi(\phi)=0$ }
\label{inf_obs_xi0}

\begin{figure}
 \centering
\subfigure[\label{rnsq1}]{\includegraphics[width=0.45\textwidth]{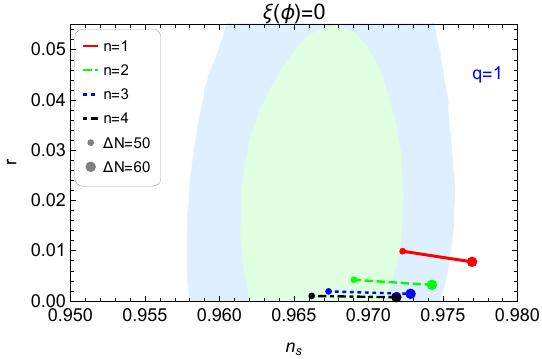}}
\subfigure[\label{rnsq2}]{\includegraphics[width=0.45\textwidth]{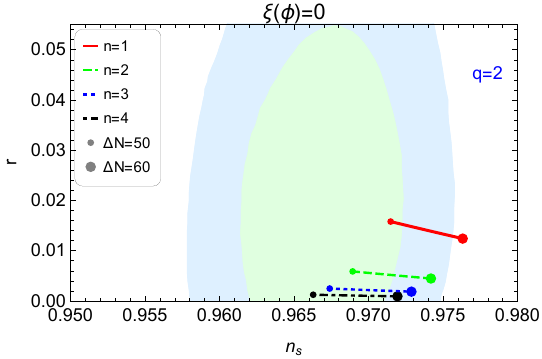}}
\subfigure[\label{rnsq3}]{\includegraphics[width=0.45\textwidth]{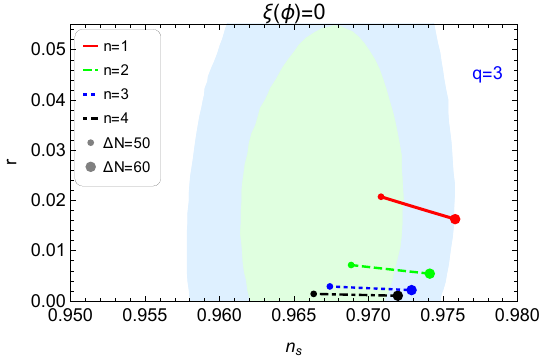}}
\subfigure[\label{rnsq4}]{\includegraphics[width=0.45\textwidth]{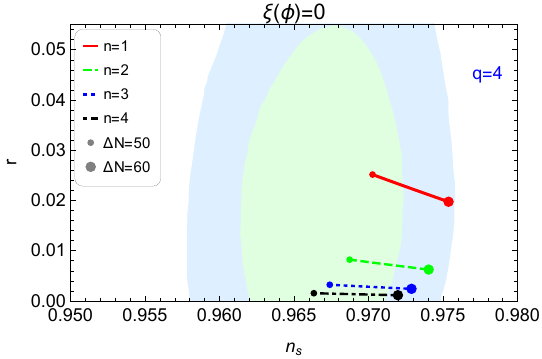}}
\caption{Plots of $r-n_s$ obtained by using the standard Slow-Roll approximation with $\xi(\phi)=0$ for different values of $n$ and $q$. The red line denotes $n=1$, the dashed green line is for $n=2$, the dotted blue line is for $n=3$, and the black dot-dashed line signifies $n=4$. The small and large dots of different colors signify $\Delta N=50$ and $\Delta N=60$ respectively. $\Delta N$ denotes the number of e-folds from the end of inflation to the onset of inflation~($\Delta N=N_{end}-N_{onset}$). The light blue shaded area represents the $2-\sigma$ bounds, while the light green shaded area represents the $1-\sigma$ bounds on $n_s$ from Planck'18 \cite{Planck:2018jri}.}
\label{rnxi0}
\end{figure}

In this section, we will present the analysis of the inflationary observables in the absence of an EGB coupling. Inserting $\xi(\phi)=0$ in Eqs. (\ref{Equ00}, \ref{Equ11}, and \ref{Equphi}) we recover equations of standard cold inflation without any contribution from the modified gravity. Following the prescription mentioned in the section (\ref{sraw})  we can obtain the slow roll parameters and inflationary observables. In Fig. \ref{rnxi0} and Table \ref{Tabrnsxi0} we have presented the analysis of $n_s$ and $r$ for different values of $n$ and $q$ for $\xi(\phi)=0$ case.

\begin{widetext}
\begin{center}
\begin{table}[h]
\begin{center}
 \resizebox{\textwidth}{!}{  
\begin{tabular}{|l| l r | l  r | l r | l r |}
\hline
\multicolumn{1}{|c}{ }&\multicolumn{1}{c}{ }&\multicolumn{1}{c}{ }&\multicolumn{1}{c}{ }&\multicolumn{1}{c}{}{$q=1$ }&\multicolumn{1}{c}{}&\multicolumn{1}{c}{}&\multicolumn{1}{c}{}&\multicolumn{1}{c|}{}\\
\hline
\multicolumn{1}{|c|}{ }&\multicolumn{1}{c}{ }{$n=1$ }&\multicolumn{1}{c|}{ }&\multicolumn{1}{c}{ }{$n=2$ }&\multicolumn{1}{c|}{}&\multicolumn{1}{c}{}{$n=3$ }&\multicolumn{1}{c|}{}&\multicolumn{1}{c}{}{$n=4$ }&\multicolumn{1}{c|}{}\\
\hline
\multicolumn{1}{|c|}{$\Delta N$}&\multicolumn{1}{c}{$r$}  &\multicolumn{1}{c|}{ $n_s$ } & \multicolumn{1}{c}{ $r$}&\multicolumn{1}{c|}{{$n_s$}}&\multicolumn{1}{c}{$r$}&\multicolumn{1}{c|}{$n_s$}&\multicolumn{1}{c}{$r$}&\multicolumn{1}{c|}{$n_s$} \\
\hline

 \,\,$50$\,&\,\,0.0098825\,\, &\,\, 0.972292\,\,&\,\,0.004194\,\,&\,\,0.969026\,\,&\,\, 0.00187688\,\,&\,\,0.967322\,\,&\,\, 0.000969976\,\,&\,\,0.966192
\\
\
 $60$\,&\,\,0.00776284\,\,&\,\,0.976953\,\,&\,\,0.00317834\,\,&\,\,0.974257\,\,&\,\,0.00139663\,\,&\,\,0.972827\,\,&\,\, 0.000713634\,\,&\,\,0.971872
\\
 \hline
\multicolumn{1}{|c}{ }&\multicolumn{1}{c}{ }&\multicolumn{1}{c}{ }&\multicolumn{1}{c}{ }&\multicolumn{1}{c}{}{$q=2$ }&\multicolumn{1}{c}{}&\multicolumn{1}{c}{}&\multicolumn{1}{c}{}&\multicolumn{1}{c|}{}\\
\hline
\multicolumn{1}{|c|}{ }&\multicolumn{1}{c}{ }{$n=1$ }&\multicolumn{1}{c|}{ }&\multicolumn{1}{c}{ }{$n=2$ }&\multicolumn{1}{c|}{}&\multicolumn{1}{c}{}{$n=3$ }&\multicolumn{1}{c|}{}&\multicolumn{1}{c}{}{$n=4$ }&\multicolumn{1}{c|}{}\\
\hline
\multicolumn{1}{|c|}{$\Delta N$}&\multicolumn{1}{c}{$r$}  &\multicolumn{1}{c|}{ $n_s$ } & \multicolumn{1}{c}{ $r$}&\multicolumn{1}{c|}{{$n_s$}}&\multicolumn{1}{c}{$r$}&\multicolumn{1}{c|}{$n_s$}&\multicolumn{1}{c}{$r$}&\multicolumn{1}{c|}{$n_s$} \\
\hline

 \,\,$50$\,&\,\,0.015774\,\, &\,\, 0.971489\,\,&\,\,0.00585307\,\,&\,\,0.96894\,\,&\,\, 0.00244552\,\,&\,\,0.9674\,\,&\,\, 0.00121017\,\,&\,\,0.966293
\\
\
 $60$\,&\,\,0.0123836\,\,&\,\,0.976325\,\,&\,\,0.00443982\,\,&\,\,0.974194\,\,&\,\,0.00182212\,\,&\,\,0.972885\,\,&\,\, 0.000891347\,\,&\,\,0.971946
\\

 \hline
 \multicolumn{1}{|c}{ }&\multicolumn{1}{c}{ }&\multicolumn{1}{c}{ }&\multicolumn{1}{c}{ }&\multicolumn{1}{c}{}{$q=3$ }&\multicolumn{1}{c}{}&\multicolumn{1}{c}{}&\multicolumn{1}{c}{}&\multicolumn{1}{c|}{}\\
\hline
\multicolumn{1}{|c|}{ }&\multicolumn{1}{c}{ }{$n=1$ }&\multicolumn{1}{c|}{ }&\multicolumn{1}{c}{ }{$n=2$ }&\multicolumn{1}{c|}{}&\multicolumn{1}{c}{}{$n=3$ }&\multicolumn{1}{c|}{}&\multicolumn{1}{c}{}{$n=4$ }&\multicolumn{1}{c|}{}\\
\hline
\multicolumn{1}{|c|}{$\Delta N$}&\multicolumn{1}{c}{$r$}  &\multicolumn{1}{c|}{ $n_s$ } & \multicolumn{1}{c}{ $r$}&\multicolumn{1}{c|}{{$n_s$}}&\multicolumn{1}{c}{$r$}&\multicolumn{1}{c|}{$n_s$}&\multicolumn{1}{c}{$r$}&\multicolumn{1}{c|}{$n_s$} \\
\hline

 \,\,$50$\,&\,\,0.0207172\,\, &\,\, 0.970843\,\,&\,\,0.00712515\,\,&\,\,0.968837\,\,&\,\, 0.00286091\,\,&\,\,0.967411\,\,&\,\, 0.00137983\,\,&\,\,0.966324
\\
\
 $60$\,&\,\,0.0162603\,\,&\,\,0.975819\,\,&\,\,0.00540743\,\,&\,\,0.974116\,\,&\,\,0.00213278\,\,&\,\,0.972894\,\,&\,\, 0.00101677\,\,&\,\,0.97197
\\
\hline
\multicolumn{1}{|c}{ }&\multicolumn{1}{c}{ }&\multicolumn{1}{c}{ }&\multicolumn{1}{c}{ }&\multicolumn{1}{c}{}{$q=4$ }&\multicolumn{1}{c}{}&\multicolumn{1}{c}{}&\multicolumn{1}{c}{}&\multicolumn{1}{c|}{}\\
\hline
\multicolumn{1}{|c|}{ }&\multicolumn{1}{c}{ }{$n=1$ }&\multicolumn{1}{c|}{ }&\multicolumn{1}{c}{ }{$n=2$ }&\multicolumn{1}{c|}{}&\multicolumn{1}{c}{}{$n=3$ }&\multicolumn{1}{c|}{}&\multicolumn{1}{c}{}{$n=4$ }&\multicolumn{1}{c|}{}\\
\hline
\multicolumn{1}{|c|}{$\Delta N$}&\multicolumn{1}{c}{$r$}  &\multicolumn{1}{c|}{ $n_s$ } & \multicolumn{1}{c}{ $r$}&\multicolumn{1}{c|}{{$n_s$}}&\multicolumn{1}{c}{$r$}&\multicolumn{1}{c|}{$n_s$}&\multicolumn{1}{c}{$r$}&\multicolumn{1}{c|}{$n_s$} \\
\hline

 \,\,$50$\,&\,\,0.0251299\,\, &\,\, 0.970275\,\,&\,\,0.00819738\,\,&\,\,0.968738\,\,&\,\, 0.0032\,\,&\,\,0.967405\,\,&\,\, 0.00151532\,\,&\,\,0.966337
\\
\
 $60$\,&\,\,0.0197209\,\,&\,\,0.975374\,\,&\,\,0.00622304\,\,&\,\,0.974041\,\,&\,\,0.00238632\,\,&\,\,0.972889\,\,&\,\, 0.00111689\,\,&\,\,0.971979
\\
\hline
\end{tabular}}
\end{center}
\caption{Table for $r-n_{s}$ for different values of potential parameters when $\xi(\phi)=0$}
\label{Tabrnsxi0}
\end{table}
\end{center}
\end{widetext}


\subsection{Numerical Calculation of Inflationary Observables }
\label{numeric_inf_obs}
\begin{figure}
\subfigure[\label{numeric_q1}]{\includegraphics[width=0.45\textwidth]{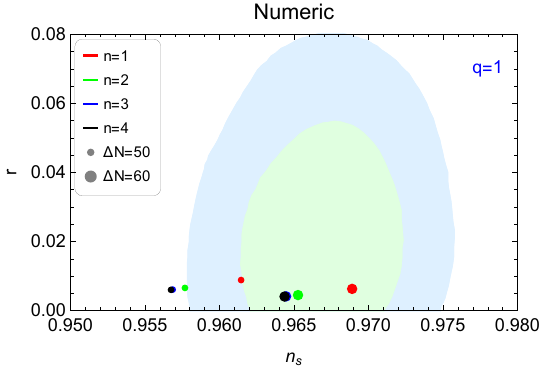}}
\subfigure[\label{numeric_q2}]{\includegraphics[width=0.45\textwidth]{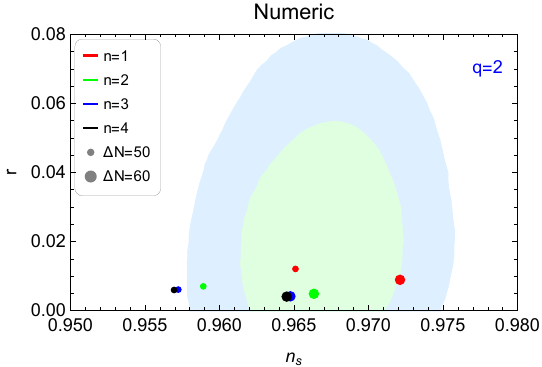}}
\subfigure[\label{numeric_q3}]{\includegraphics[width=0.45\textwidth]{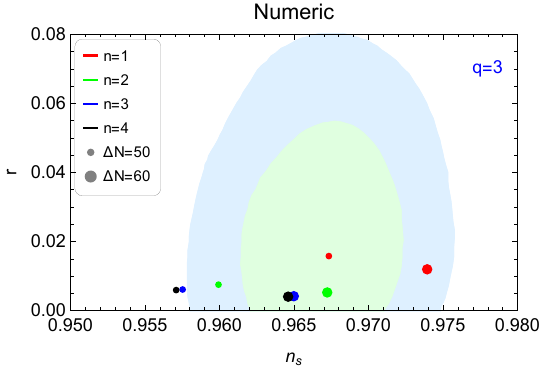}}
\subfigure[\label{numeric_q4}]{\includegraphics[width=0.45\textwidth]{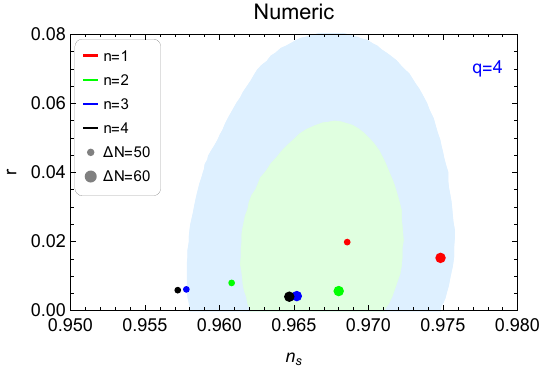}}
\caption{Plots of $r-n_s$ obtained by  solving system (~\ref{DynSYSN}) numerically for different values of $n$ and $q$. The red dot denotes the $n=1$, the green dot is for $n=2$, the blue dot is for $n=3$, and the black dot signifies $n=4$. The small and large dots of different colors signify the $\Delta N=50$ and $\Delta N=60$, respectively. $\Delta N$ denotes the number of e-folds from the end of inflation to the onset of inflation~($\Delta N=N_{end}-N_{onset}$).  The light blue shaded area is for the $2-\sigma$ bounds, while the light green shaded area represents the $1-\sigma$ bounds on $n_s$ from Planck'18 \cite{Planck:2018jri}} 
\label{rns_numeric}
\end{figure}

In this section, we present the numerical calculation of the inflationary observables. We obtain the numerical solutions for  $\phi, \chi$ and $Q$ in terms of the number of e-folds ($N$) by solving the system~ (\ref{DynSYSN}). To obtain the solution, we need to solve the system~ (\ref{DynSYSN}) numerically, which requires the initial condition to start with. We choose a value of the initial $\phi$>0 and $\chi< 0$, for $\chi< 0$, the inflationary trajectory is an attractor.  Inserting the value of $\phi$ and $\chi$ into the Eq. \ref{Equ00N} we compute the initial value of Q, after obtaining the required initial conditions we can easily solve the system(\ref{DynSYSN}) by evolving the system for a range of e-folds ($N$). Once we obtained the numerical solutions, it is straightforward to compute the slow roll parameters using Eq. (\ref{epsilon}) and (\ref{delta}). Inserting the slow roll parameters in Eq.~(\ref{ns_slr}) and (\ref{r_slr}), we calculate the scalar spectral index ($n_s$) and tensor to scalar ratio ($r$). The values of $n_s$ and $r$ obtained here are exact without using any approximations. (see Fig. \ref{rns_numeric}). The exact numerical values are in Table (\ref{tab_rns_numeric}). We take $\xi_1=10$ and $\xi_2 =0.4$, these values of the coupling parameters are chosen so that the constraints on $n_s$ and $r$ do not contradict with the current CMB bounds.

\begin{figure}
\subfigure[\label{rnsslow1q1}]{\includegraphics[width=0.45\textwidth]{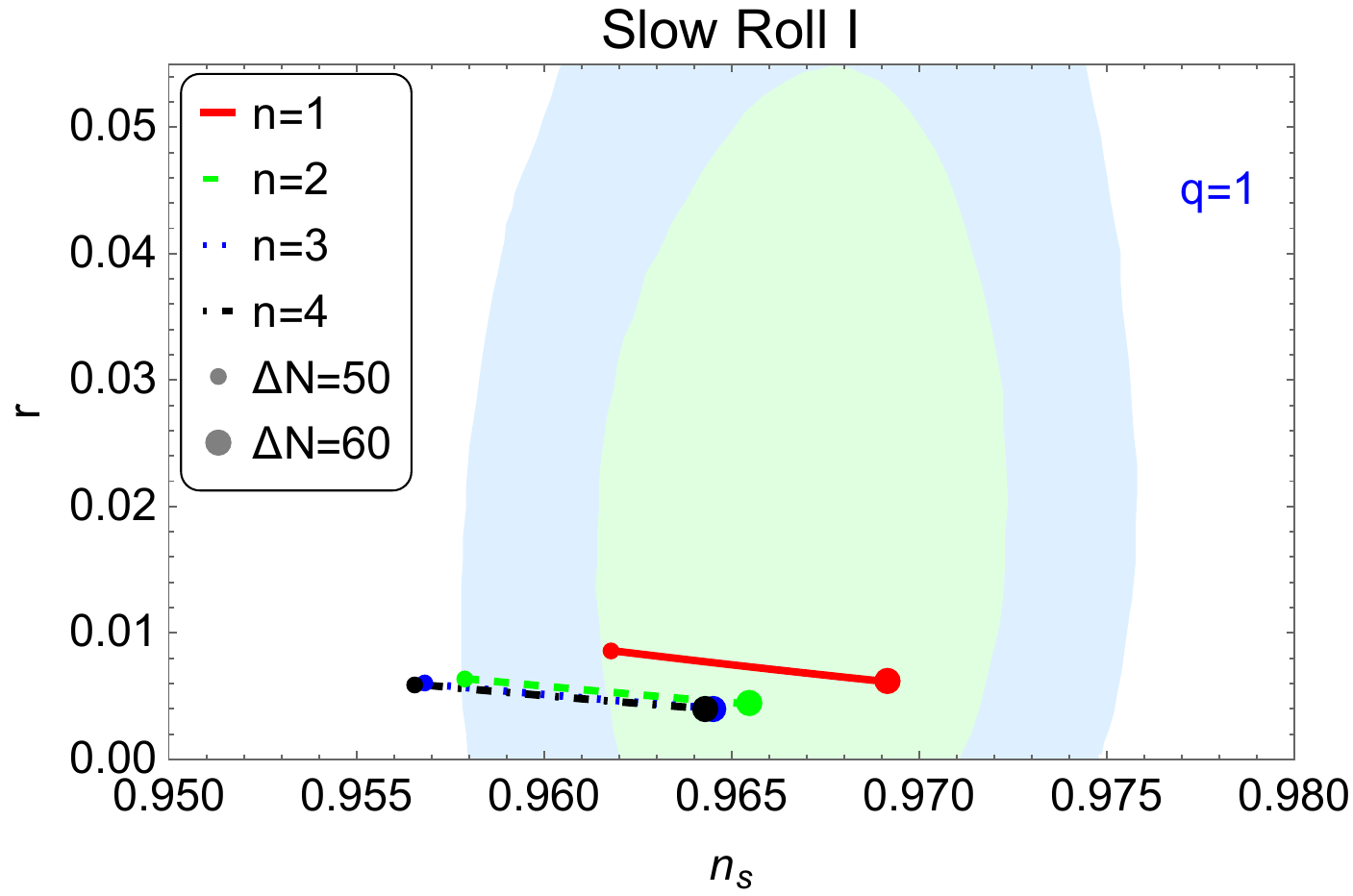}}
\subfigure[\label{rnsslow1q2}]{\includegraphics[width=0.45\textwidth]{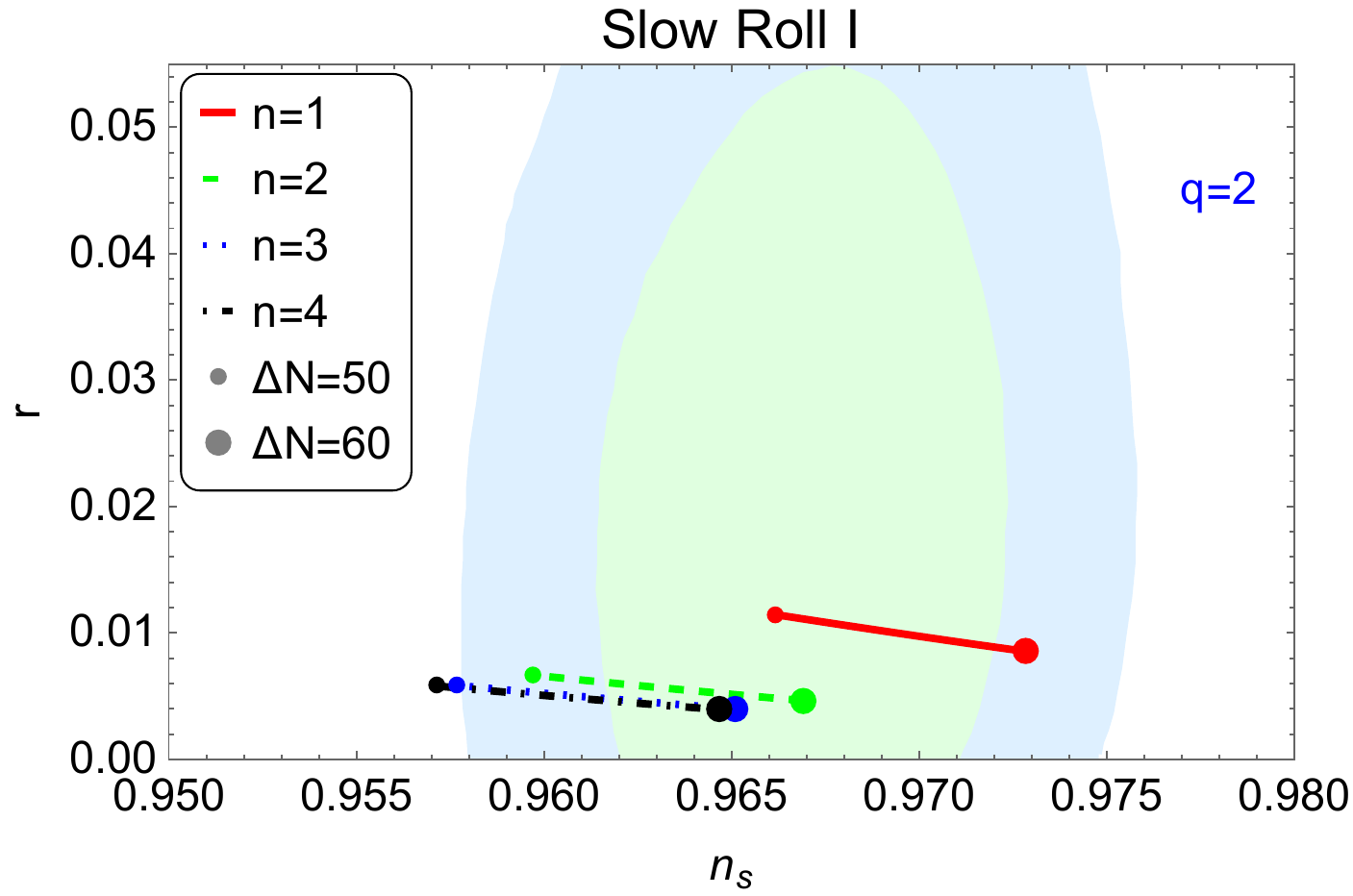}}
\subfigure[\label{rnsslow1q3}]{\includegraphics[width=0.45\textwidth]{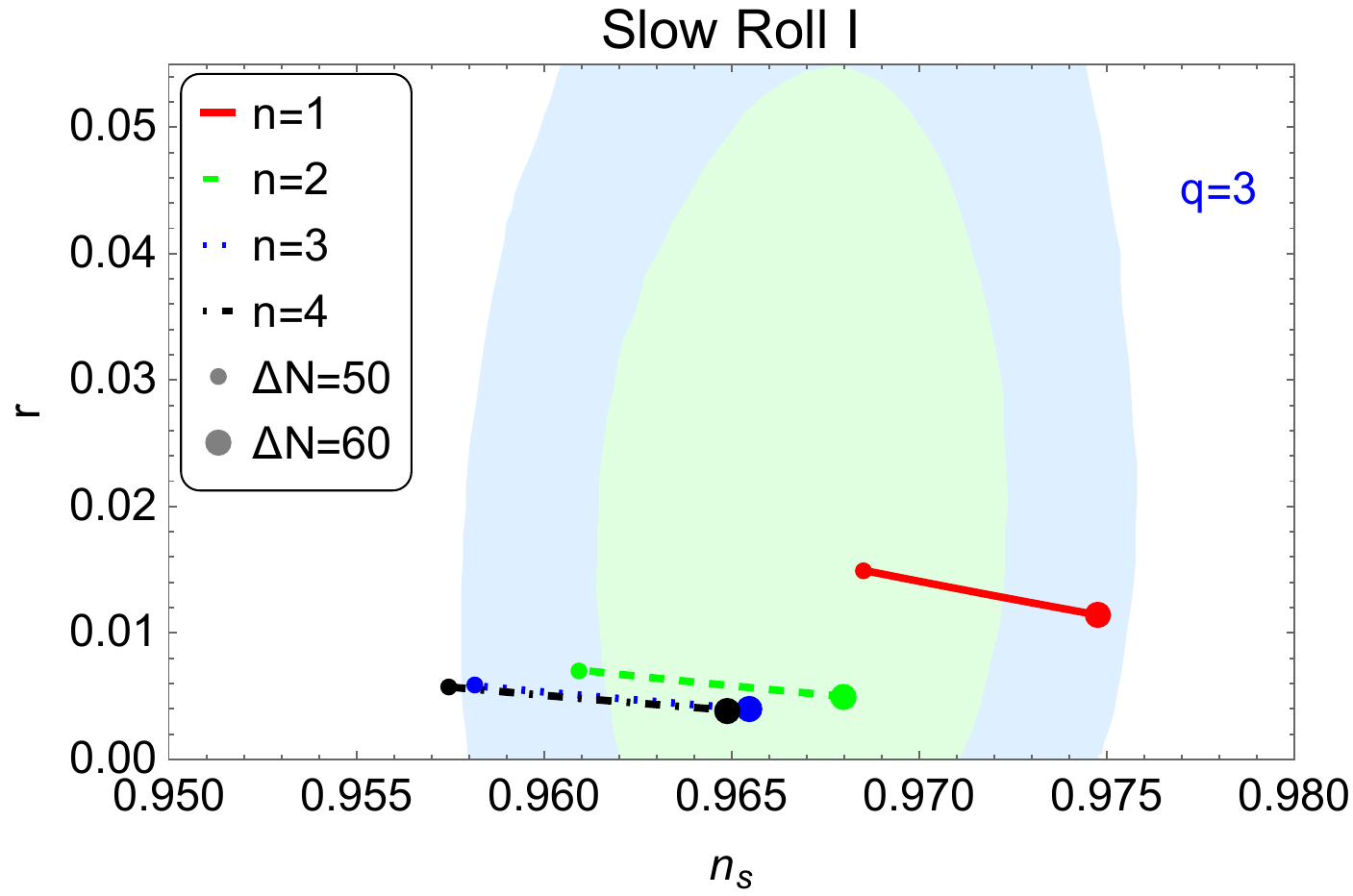}}
\subfigure[\label{rnsslow1q4}]{\includegraphics[width=0.45\textwidth]{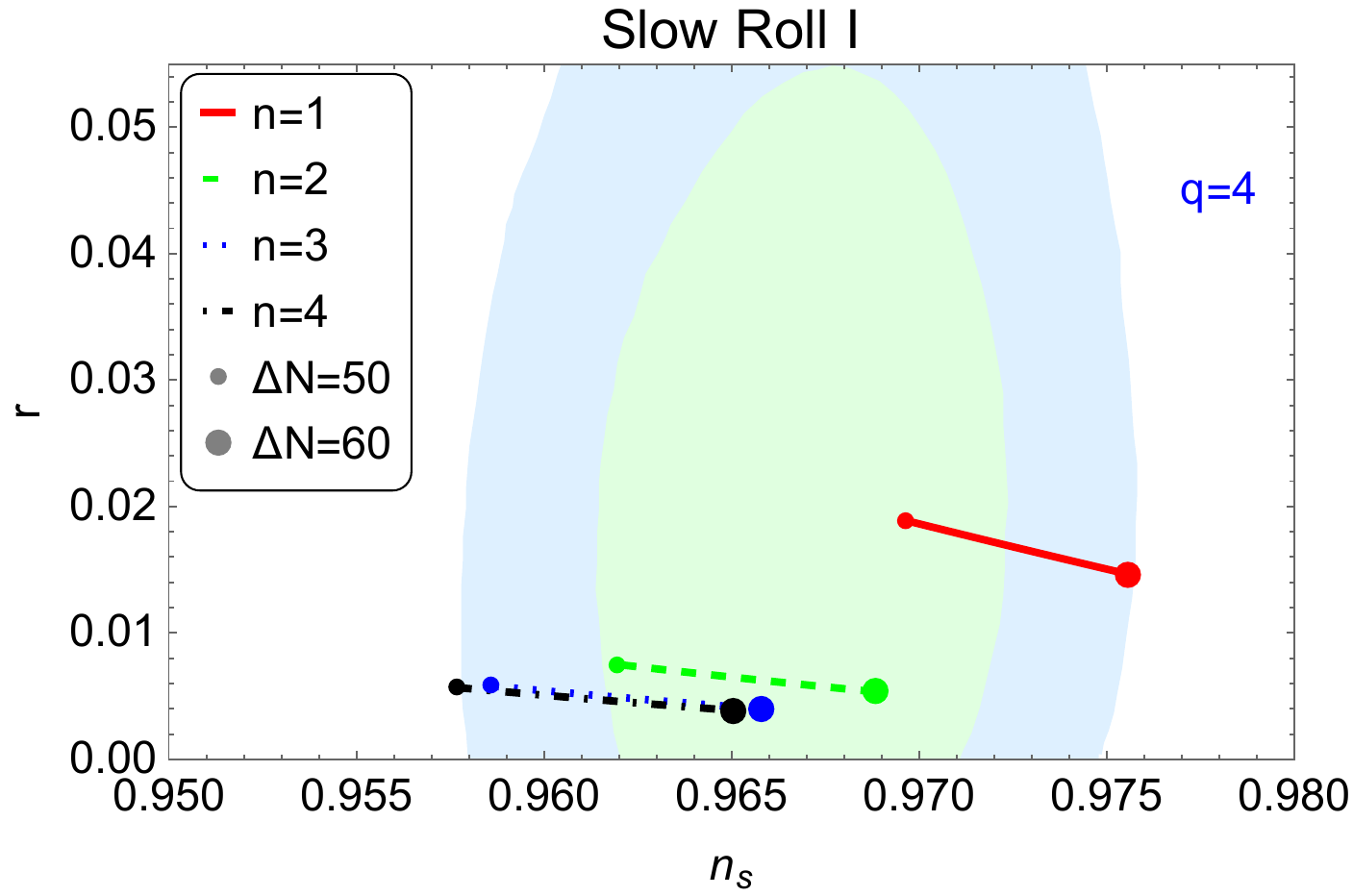}}
\caption{Plots of $r-n_s$ obtained by using Slow-Roll Approximation I for different values of $n$ and $q$. The red line denotes $n=1$, the dashed green line is for $n=2$, the dotted blue line is for $n=3$, and the black dot-dashed line signifies $n=4$. The small and large dots of different colors signify $\Delta N=50$ and $\Delta N=60$ respectively. $\Delta N$ denotes the number of e-folds from the end of inflation to the onset of inflation~($\Delta N=N_{end}-N_{onset}$).  The light blue shaded area represents the $2-\sigma$ bounds, while the light green shaded area represents the $1-\sigma$ bounds on $n_s$ from Planck'18 \cite{Planck:2018jri}} 
\label{rnsslow1}
\end{figure}


\begin{widetext}
\begin{center}
\begin{table}[h]
\begin{center}
 \resizebox{\textwidth}{!}{  
\begin{tabular}{|l| l r | l  r | l r | l r |}
\hline
\multicolumn{1}{|c}{ }&\multicolumn{1}{c}{ }&\multicolumn{1}{c}{ }&\multicolumn{1}{c}{ }&\multicolumn{1}{c}{}{{{$q=1$} }}&\multicolumn{1}{c}{}&\multicolumn{1}{c}{}&\multicolumn{1}{c}{}&\multicolumn{1}{c|}{}\\
\hline
\multicolumn{1}{|c|}{ }&\multicolumn{1}{c}{ }{$n=1$ }&\multicolumn{1}{c|}{ }&\multicolumn{1}{c}{ }{$n=2$ }&\multicolumn{1}{c|}{}&\multicolumn{1}{c}{}{$n=3$ }&\multicolumn{1}{c|}{}&\multicolumn{1}{c}{}{$n=4$ }&\multicolumn{1}{c|}{}\\
\hline
\multicolumn{1}{|c|}{$\Delta N$}&\multicolumn{1}{c}{$r$}  &\multicolumn{1}{c|}{ $n_s$ } & \multicolumn{1}{c}{ $r$}&\multicolumn{1}{c|}{{$n_s$}}&\multicolumn{1}{c}{$r$}&\multicolumn{1}{c|}{$n_s$}&\multicolumn{1}{c}{$r$}&\multicolumn{1}{c|}{$n_s$} \\
\hline

 \,\,$50$\,&\,\,0.00881346\,\, &\,\, 0.961455\,\,&\,\,0.00653732\,\,&\,\,0.957687\,\,&\,\, 0.00604871\,\,&\,\,0.956863\,\,&\,\, 0.00597986\,\,&\,\,0.956753
\\
\ 
 $60$\,&\,\,0.00626066\,\,&\,\,0.968906\,\,&\,\,0.00446308\,\,&\,\,0.965274\,\,&\,\,0.00409352\,\,&\,\,0.964475\,\,&\,\, 0.00404247\,\,&\,\,0.96437
\\
 \hline
\multicolumn{1}{|c}{ }&\multicolumn{1}{c}{ }&\multicolumn{1}{c}{ }&\multicolumn{1}{c}{ }&\multicolumn{1}{c}{}{{{$q=2$} } }&\multicolumn{1}{c}{}&\multicolumn{1}{c}{}&\multicolumn{1}{c}{}&\multicolumn{1}{c|}{}\\
\hline
\multicolumn{1}{|c|}{ }&\multicolumn{1}{c}{ }{$n=1$ }&\multicolumn{1}{c|}{ }&\multicolumn{1}{c}{ }{$n=2$ }&\multicolumn{1}{c|}{}&\multicolumn{1}{c}{}{$n=3$ }&\multicolumn{1}{c|}{}&\multicolumn{1}{c}{}{$n=4$ }&\multicolumn{1}{c|}{}\\
\hline
\multicolumn{1}{|c|}{$\Delta N$}&\multicolumn{1}{c}{$r$}  &\multicolumn{1}{c|}{ $n_s$ } & \multicolumn{1}{c}{ $r$}&\multicolumn{1}{c|}{{$n_s$}}&\multicolumn{1}{c}{$r$}&\multicolumn{1}{c|}{$n_s$}&\multicolumn{1}{c}{$r$}&\multicolumn{1}{c|}{$n_s$} \\
\hline

 \,\,$50$\,&\,\,0.0120394\,\, &\,\, 0.965104\,\,&\,\,0.00699186\,\,&\,\,0.958916\,\,&\,\, 0.00605133\,\,&\,\,0.957241\,\,&\,\, 0.00593393\,\,&\,\,0.956958
\\
\
 $60$\,&\,\,0.00888793\,\,&\,\,0.972119\,\,&\,\,0.00483255\,\,&\,\,0.966341\,\,&\,\,0.00410949\,\,&\,\,0.964764\,\,&\,\, 0.00401795\,\,&\,\,0.964514
\\

 \hline
 \multicolumn{1}{|c}{ }&\multicolumn{1}{c}{ }&\multicolumn{1}{c}{ }&\multicolumn{1}{c}{ }&\multicolumn{1}{c}{}{{{$q=3$} } }&\multicolumn{1}{c}{}&\multicolumn{1}{c}{}&\multicolumn{1}{c}{}&\multicolumn{1}{c|}{}\\
\hline
\multicolumn{1}{|c|}{ }&\multicolumn{1}{c}{ }{$n=1$ }&\multicolumn{1}{c|}{ }&\multicolumn{1}{c}{ }{$n=2$ }&\multicolumn{1}{c|}{}&\multicolumn{1}{c}{}{$n=3$ }&\multicolumn{1}{c|}{}&\multicolumn{1}{c}{}{$n=4$ }&\multicolumn{1}{c|}{}\\
\hline
\multicolumn{1}{|c|}{$\Delta N$}&\multicolumn{1}{c}{$r$}  &\multicolumn{1}{c|}{ $n_s$ } & \multicolumn{1}{c}{ $r$}&\multicolumn{1}{c|}{{$n_s$}}&\multicolumn{1}{c}{$r$}&\multicolumn{1}{c|}{$n_s$}&\multicolumn{1}{c}{$r$}&\multicolumn{1}{c|}{$n_s$} \\
\hline

 \,\,$50$\,&\,\,0.0157547\,\, &\,\, 0.967343\,\,&\,\,0.00748015\,\,&\,\,0.959938\,\,&\,\, 0.00607689\,\,&\,\,0.957532\,\,&\,\, 0.00590639\,\,&\,\,0.957097
\\
\
 $60$\,&\,\,0.0119274\,\,&\,\,0.973941\,\,&\,\,0.00522369\,\,&\,\,0.967235\,\,&\,\,0.0041382\,\,&\,\,0.964991\,\,&\,\, 0.0040042\,\,&\,\,0.964611
\\
\hline
\multicolumn{1}{|c}{ }&\multicolumn{1}{c}{ }&\multicolumn{1}{c}{ }&\multicolumn{1}{c}{ }&\multicolumn{1}{c}{}{{{$q=4$} } }&\multicolumn{1}{c}{}&\multicolumn{1}{c}{}&\multicolumn{1}{c}{}&\multicolumn{1}{c|}{}\\
\hline
\multicolumn{1}{|c|}{ }&\multicolumn{1}{c}{ }{$n=1$ }&\multicolumn{1}{c|}{ }&\multicolumn{1}{c}{ }{$n=2$ }&\multicolumn{1}{c|}{}&\multicolumn{1}{c}{}{$n=3$ }&\multicolumn{1}{c|}{}&\multicolumn{1}{c}{}{$n=4$ }&\multicolumn{1}{c|}{}\\
\hline
\multicolumn{1}{|c|}{$\Delta N$}&\multicolumn{1}{c}{$r$}  &\multicolumn{1}{c|}{ $n_s$ } & \multicolumn{1}{c}{ $r$}&\multicolumn{1}{c|}{{$n_s$}}&\multicolumn{1}{c}{$r$}&\multicolumn{1}{c|}{$n_s$}&\multicolumn{1}{c}{$r$}&\multicolumn{1}{c|}{$n_s$} \\
\hline

 \,\,$50$\,&\,\,0.0198034\,\, &\,\, 0.968575\,\,&\,\,0.00798492\,\,&\,\,0.96082\,\,&\,\, 0.00611248\,\,&\,\,0.957782\,\,&\,\, 0.00588756\,\,&\,\,0.957205
\\
\
 $60$\,&\,\,0.0152318\,\,&\,\,0.974837\,\,&\,\,0.00562672\,\,&\,\,0.968004\,\,&\,\,0.00417223\,\,&\,\,0.965191\,\,&\,\, 0.00399504\,\,&\,\,0.964688
\\
\hline
\end{tabular}}
\end{center}
\caption{Table for $r-n_{s}$ for different values of potential parameters by solving system (~\ref{DynSYSN}) numerically for different values of $n$ and $q$}
\label{tab_rns_numeric}
\end{table}
\end{center}
\end{widetext}

\subsection{Inflationary Observables in Slow-Roll I}
 The knowledge of the slow-roll parameters as functions of $\phi$ allows us to get $n_s(\phi)$, $r(\phi)$, and $A_s(\phi)$  using formulae (\ref{ns_slr})--(\ref{As_slr}).
Now, one can implement the above findings for the specific case of the PLP model. One can get the analytic expressions for the slow-roll parameters using the new approximation-I as given in \ref{App1}.
\begin{figure}
\subfigure[\label{rnsslow2q1}]{\includegraphics[width=0.45\textwidth]{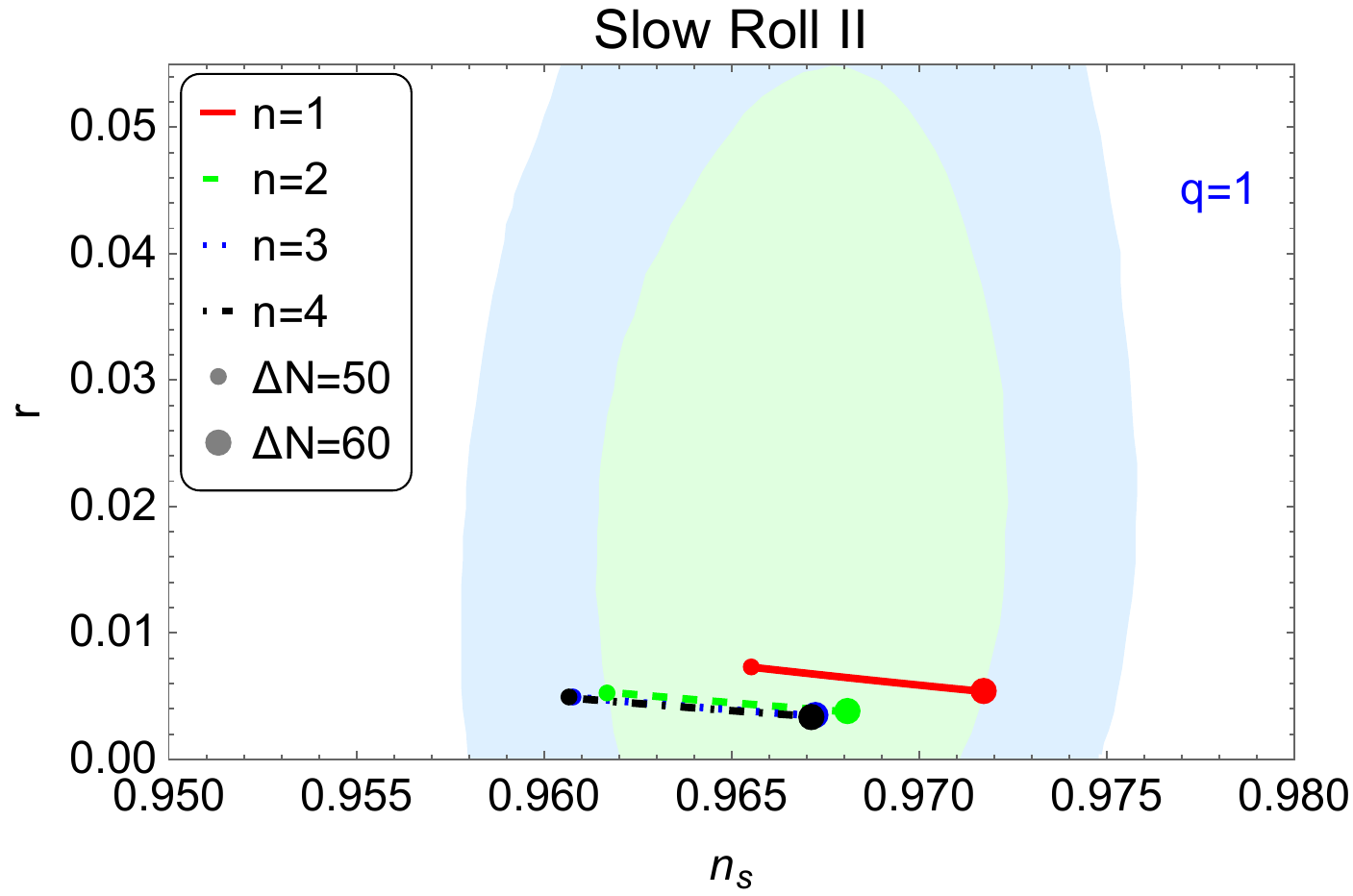}}
\subfigure[\label{rnsslow2q2}]{\includegraphics[width=0.45\textwidth]{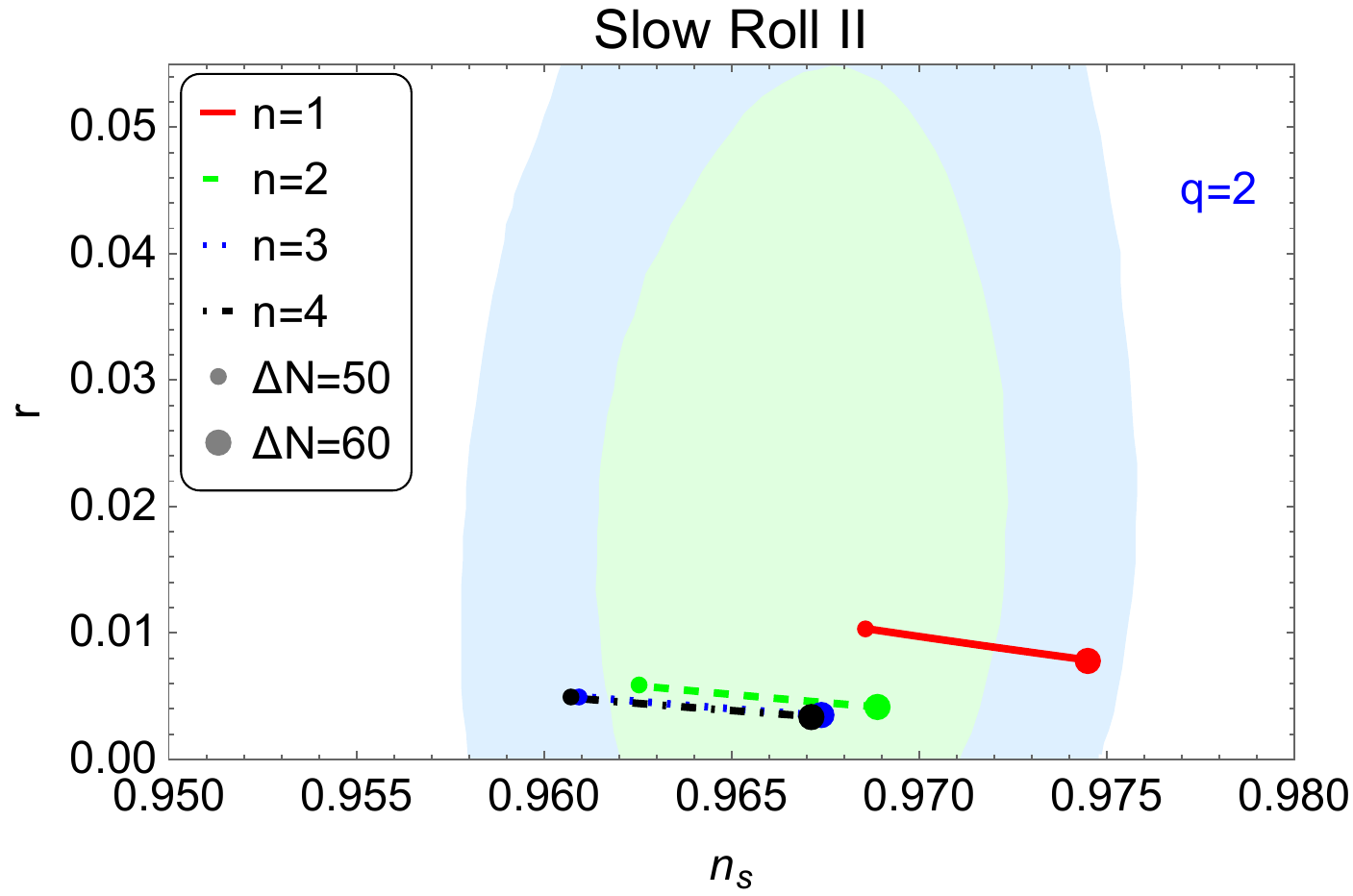}}
\subfigure[\label{rnsslow2q3}]{\includegraphics[width=0.45\textwidth]{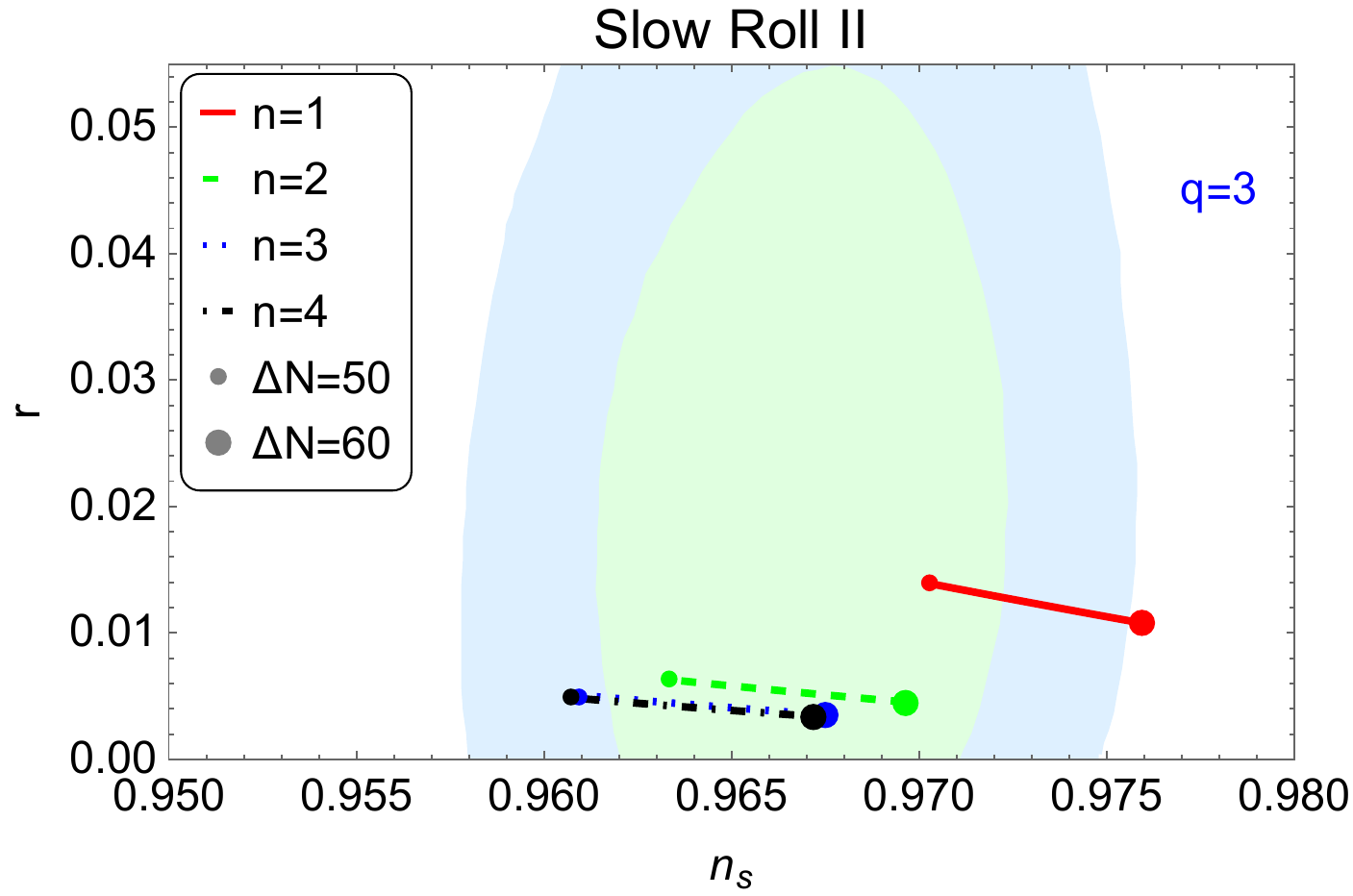}}
\subfigure[\label{rnsslow2q4}]{\includegraphics[width=0.45\textwidth]{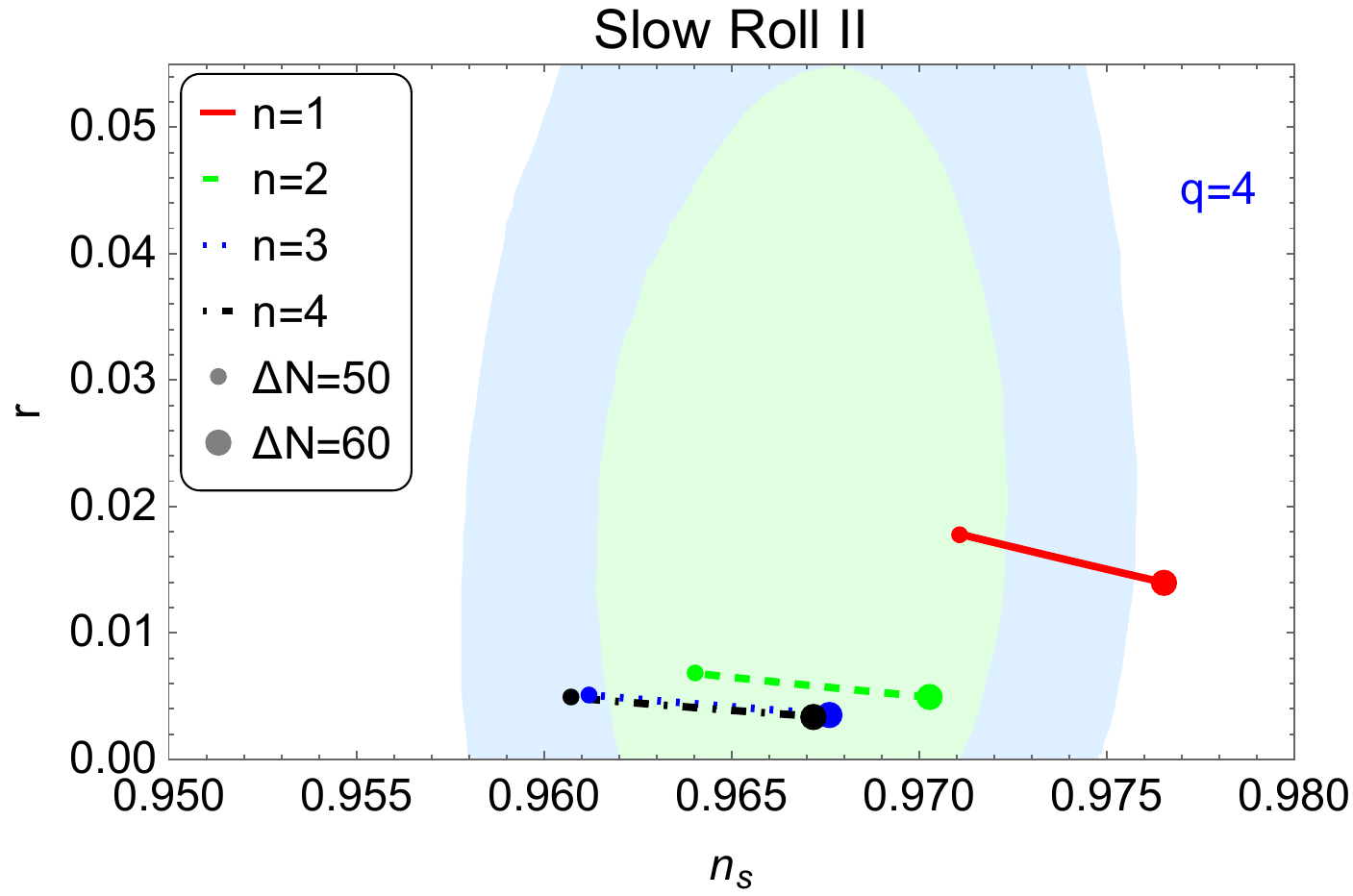}}
\caption{Plots of $r-n_s$ obtained by using Slow-Roll Approximation II for different values of $n$ and $q$. The red line denotes $n=1$, the dashed green line is for $n=2$, the dotted blue line is for $n=3$, and the black dot-dashed line signifies $n=4$. The small and large dots of different colors signify $\Delta N=50$ and $\Delta N=60$, respectively. $\Delta N$ denotes the number of e-folds from the end of inflation to the onset of inflation~($\Delta N=N_{end}-N_{onset}$).  The light blue shaded area represents the $2-\sigma$ bounds, while the light green shaded area represents the $1-\sigma$ bounds on $n_s$ from Planck'18 \cite{Planck:2018jri}} 
\label{rnsslow2}
\end{figure}


Of course, the results look cumbersome, to say the least, but one should appreciate the fact that using these techniques, one can still obtain analytical solutions. 

Now, we will discuss the PLP inflation in slow-roll approximation I. To calculate the inflationary observables, first we calculate the slow roll parameters, using Eq.(~\ref{apprIeps1phi}) and Eq.(~\ref{delta1phi})we compute $\varepsilon_1, \delta_1$. It is then straightforward to obtain $\varepsilon_2$ and $\delta_2$ using Eq.(~\ref{eps2delta2phi}). When studying inflation, another important quantity is the number of e-folds $N$. Thus, from Eq.(~\ref{apprIdNDdphi}), one can write $N$ for a given $\phi$. To solve Eq.(~\ref{apprIdNDdphi}), we followed the same prescription as mentioned in \cite{Pozdeeva:2024ihc}, and we have taken $N=60$ at the end of inflation, whereas $N=0$ corresponds to the onset of inflation.

We set the initial field value for solving  Eq.(~\ref{apprIdNDdphi}) at the end of inflation, which can be calculated from $\varepsilon_1=1$. For a particular value of $n =(1-4)$ and $q= (1-4)$ with $\xi_1=10, \xi_2=0.4$ we can calculate the field value where $\varepsilon=1$ and we use this as the end of inflation ($\phi_{e}$). We repeat this for all the possible combinations of $n$ and $q$ after acquiring the values of the $\phi_e$ for combinations of $n$ and $q$. We use $\phi_{e}$ as the initial condition to solve Eq.(\ref{apprIdNDdphi}). It is difficult to solve Eq.(\ref{apprIdNDdphi}) analytically, so we then employ numerical techniques. Equipped with all the preliminary equations in hand, one can finally compute the scalar spectral index ($n_s$), the tensor to the scalar ratio ($r$), and the amplitude of scalar power spectrum ($A_s$) using  Eq.(\ref{ns_slr}),(\ref{r_slr}) and Eq.(\ref{As_slr}) respectively. We keep the same values of coupling parameters  $\xi_1=10$ and $\xi_2 =0.4$. Whereas, $V_0$ can be fixed by matching $A_s=2.09\times10^{-9}$ at the pivot scale. 

 As we have previously mentioned $n$ and $q$ vary from $1$ to $4$, which results in the $16$ different classes of potential. We have analyzed all the possible combinations. 
In fig. \ref{rnsslow1} we present the results on the $r-n_s$ where Slow-Roll I approximation has been utilized. It is evident from fig.  \ref{rnsslow1} that almost all the obtained values of $r$ and $n_s$ for combination of potential parameters $n$ and $q$ are well inside the constraints imposed by the {\it Planck 18}. Explicit numerical values of the inflationary observables can be found in Table \ref{Table1}. 

\begin{widetext}
\begin{center}
\begin{table}[h]
\begin{center}
 \resizebox{\textwidth}{!}{  
\begin{tabular}{|l| l r | l  r | l r | l r |}
\hline
\multicolumn{1}{|c}{ }&\multicolumn{1}{c}{ }&\multicolumn{1}{c}{ }&\multicolumn{1}{c}{ }&\multicolumn{1}{c}{}{$q=1$ }&\multicolumn{1}{c}{}&\multicolumn{1}{c}{}&\multicolumn{1}{c}{}&\multicolumn{1}{c|}{}\\
\hline
\multicolumn{1}{|c|}{ }&\multicolumn{1}{c}{ }{$n=1$ }&\multicolumn{1}{c|}{ }&\multicolumn{1}{c}{ }{$n=2$ }&\multicolumn{1}{c|}{}&\multicolumn{1}{c}{}{$n=3$ }&\multicolumn{1}{c|}{}&\multicolumn{1}{c}{}{$n=4$ }&\multicolumn{1}{c|}{}\\
\hline
\multicolumn{1}{|c|}{$\Delta N$}&\multicolumn{1}{c}{$r$}  &\multicolumn{1}{c|}{ $n_s$ } & \multicolumn{1}{c}{ $r$}&\multicolumn{1}{c|}{{$n_s$}}&\multicolumn{1}{c}{$r$}&\multicolumn{1}{c|}{$n_s$}&\multicolumn{1}{c}{$r$}&\multicolumn{1}{c|}{$n_s$} \\
\hline

 \,\,$50$\,&\,\,0.008604\,\, &\,\, 0.961780\,\,&\,\,0.006397\,\,&\,\,0.957867\,\,&\,\, 0.005970\,\,&\,\,0.9568\,\,&\,\, 0.0059473\,\,&\,\,0.956564
\\
\
 $60$\,&\,\,0.006130\,\,&\,\,0.969175\,\,&\,\,0.0043752\,\,&\,\,0.96545\,\,&\,\,0.0059704\,\,&\,\,0.96450\,\,&\,\, 0.00401488\,\,&\,\,0.96429
\\
 \hline
\multicolumn{1}{|c}{ }&\multicolumn{1}{c}{ }&\multicolumn{1}{c}{ }&\multicolumn{1}{c}{ }&\multicolumn{1}{c}{}{$q=2$ }&\multicolumn{1}{c}{}&\multicolumn{1}{c}{}&\multicolumn{1}{c}{}&\multicolumn{1}{c|}{}\\
\hline
\multicolumn{1}{|c|}{ }&\multicolumn{1}{c}{ }{$n=1$ }&\multicolumn{1}{c|}{ }&\multicolumn{1}{c}{ }{$n=2$ }&\multicolumn{1}{c|}{}&\multicolumn{1}{c}{}{$n=3$ }&\multicolumn{1}{c|}{}&\multicolumn{1}{c}{}{$n=4$ }&\multicolumn{1}{c|}{}\\
\hline
\multicolumn{1}{|c|}{$\Delta N$}&\multicolumn{1}{c}{$r$}  &\multicolumn{1}{c|}{ $n_s$ } & \multicolumn{1}{c}{ $r$}&\multicolumn{1}{c|}{{$n_s$}}&\multicolumn{1}{c}{$r$}&\multicolumn{1}{c|}{$n_s$}&\multicolumn{1}{c}{$r$}&\multicolumn{1}{c|}{$n_s$} \\
\hline

 \,\,$50$\,&\,\,0.0114495\,\, &\,\, 0.96616\,\,&\,\,0.006667\,\,&\,\,0.959685\,\,&\,\, 0.0058103\,\,&\,\,0.957663\,\,&\,\, 0.00580653\,\,&\,\,0.95712
\\
\
 $60$\,&\,\,0.008526\,\,&\,\,0.972857\,\,&\,\,0.004638\,\,&\,\,0.966922\,\,&\,\,0.00398821\,\,&\,\,0.965102\,\,&\,\, 0.00393797\,\,&\,\,0.964671
\\

 \hline
 \multicolumn{1}{|c}{ }&\multicolumn{1}{c}{ }&\multicolumn{1}{c}{ }&\multicolumn{1}{c}{ }&\multicolumn{1}{c}{}{$q=3$ }&\multicolumn{1}{c}{}&\multicolumn{1}{c}{}&\multicolumn{1}{c}{}&\multicolumn{1}{c|}{}\\
\hline
\multicolumn{1}{|c|}{ }&\multicolumn{1}{c}{ }{$n=1$ }&\multicolumn{1}{c|}{ }&\multicolumn{1}{c}{ }{$n=2$ }&\multicolumn{1}{c|}{}&\multicolumn{1}{c}{}{$n=3$ }&\multicolumn{1}{c|}{}&\multicolumn{1}{c}{}{$n=4$ }&\multicolumn{1}{c|}{}\\
\hline
\multicolumn{1}{|c|}{$\Delta N$}&\multicolumn{1}{c}{$r$}  &\multicolumn{1}{c|}{ $n_s$ } & \multicolumn{1}{c}{ $r$}&\multicolumn{1}{c|}{{$n_s$}}&\multicolumn{1}{c}{$r$}&\multicolumn{1}{c|}{$n_s$}&\multicolumn{1}{c}{$r$}&\multicolumn{1}{c|}{$n_s$} \\
\hline

 \,\,$50$\,&\,\,0.0149482\,\, &\,\, 0.968519\,\,&\,\,0.007059\,\,&\,\,0.960959\,\,&\,\, 0.00581401\,\,&\,\,0.958181\,\,&\,\, 0.00572878\,\,&\,\,0.957447
\\
\
 $60$\,&\,\,0.0114241\,\,&\,\,0.974752\,\,&\,\,0.004972\,\,&\,\,0.967989\,\,&\,\,0.00398117\,\,&\,\,0.965487\,\,&\,\, 0.00389589\,\,&\,\,0.964897
\\
\hline
\multicolumn{1}{|c}{ }&\multicolumn{1}{c}{ }&\multicolumn{1}{c}{ }&\multicolumn{1}{c}{ }&\multicolumn{1}{c}{}{$q=4$ }&\multicolumn{1}{c}{}&\multicolumn{1}{c}{}&\multicolumn{1}{c}{}&\multicolumn{1}{c|}{}\\
\hline
\multicolumn{1}{|c|}{ }&\multicolumn{1}{c}{ }{$n=1$ }&\multicolumn{1}{c|}{ }&\multicolumn{1}{c}{ }{$n=2$ }&\multicolumn{1}{c|}{}&\multicolumn{1}{c}{}{$n=3$ }&\multicolumn{1}{c|}{}&\multicolumn{1}{c}{}{$n=4$ }&\multicolumn{1}{c|}{}\\
\hline
\multicolumn{1}{|c|}{$\Delta N$}&\multicolumn{1}{c}{$r$}  &\multicolumn{1}{c|}{ $n_s$ } & \multicolumn{1}{c}{ $r$}&\multicolumn{1}{c|}{{$n_s$}}&\multicolumn{1}{c}{$r$}&\multicolumn{1}{c|}{$n_s$}&\multicolumn{1}{c}{$r$}&\multicolumn{1}{c|}{$n_s$} \\
\hline

 \,\,$50$\,&\,\,0.0188761\,\, &\,\, 0.969662\,\,&\,\,0.00750292\,\,&\,\,0.961960\,\,&\,\, 0.0058097\,\,&\,\,0.958572\,\,&\,\, 0.0056763\,\,&\,\,0.956782
\\
\
 $60$\,&\,\,0.0146428\,\,&\,\,0.975524\,\,&\,\,0.00533771\,\,&\,\,0.968838\,\,&\,\,0.003992\,\,&\,\,0.965786\,\,&\,\, 0.00386787\,\,&\,\,0.965061
\\
\hline
\end{tabular}}
\end{center}
\caption{Table for $r-n_{s}$ for different values of potential parameters in Slow-Roll I}
\label{Table1}
\end{table}
\end{center}
\end{widetext}


\subsection{Inflationary Observables in Slow-Roll II}

    Using the obtained expressions of $\delta_1(\phi)$ and $\varepsilon_1(\phi)$ and Eq.(~\ref{eps2delta2phi}), one can now calculate parameters $\varepsilon_2(\phi)$ and $\delta_2(\phi)$.
One can again get the analytic expressions for the slow-roll parameters using the new approximation-II as given in \ref{app2}. After this, one can, in the same spirit, obtain $n_s(\phi)$, $r(\phi)$, and $A_s(\phi)$  using Eq.~(\ref{ns_slr})--(\ref{As_slr}).

\begin{figure}
    \centering
    \includegraphics[width=0.5\linewidth]{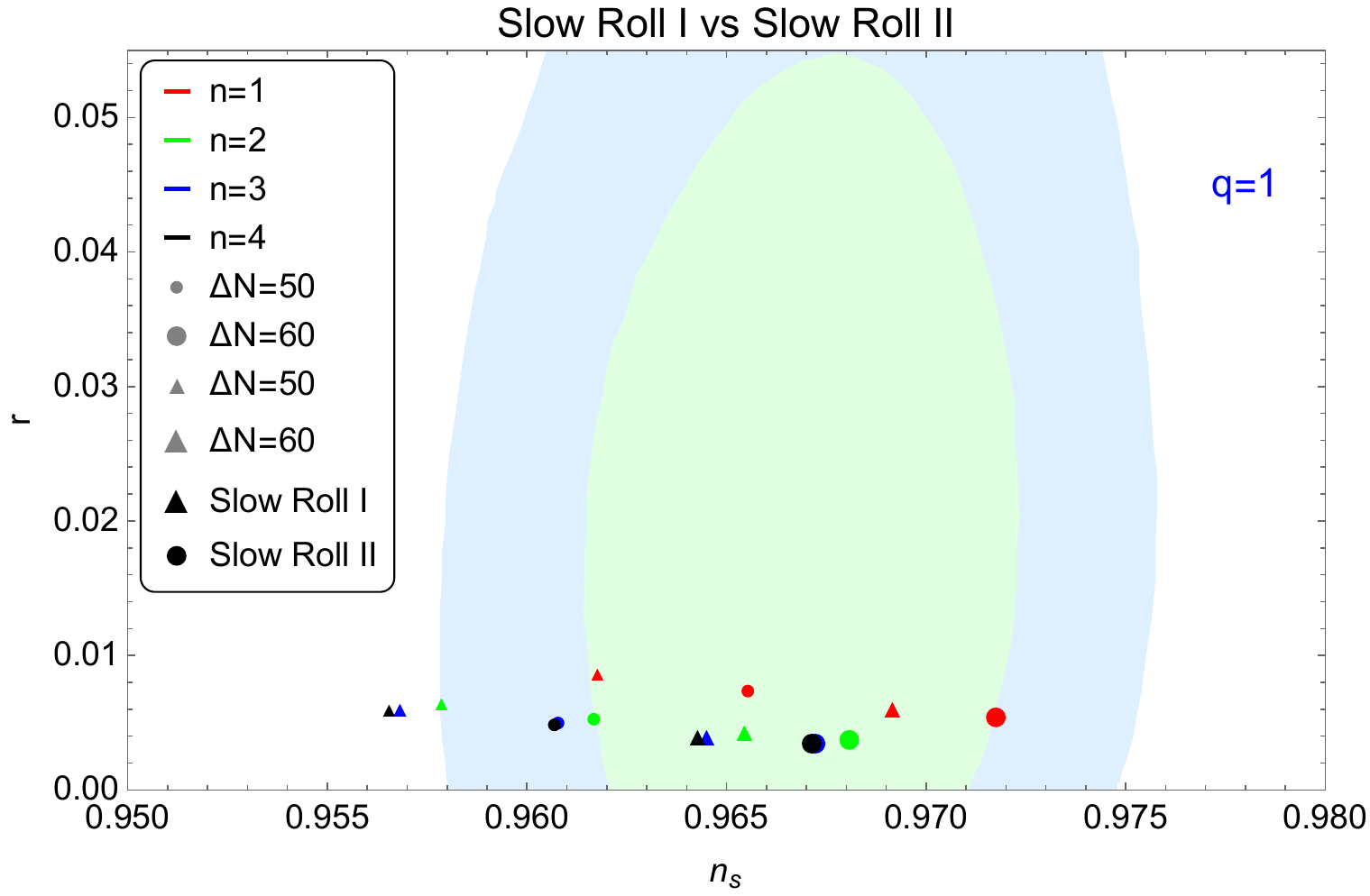}
    \caption{The comparison between the Slow Roll I and Slow Roll II for the PLP potential with $ q=1$ and $ n=1-4$. The legends are self-explanatory.}
    \label{slow1slow2}
\end{figure}

In this section, we will discuss the PLP inflation in slow-roll approximation II.  We employ the same method to calculate the inflationary observables as we did in Slow Roll Approximation I.  First, we calculate the slow roll parameters ($\varepsilon_1, \varepsilon_2, \delta_1,\delta_2$), using Eq.(~\ref{apprIIeps1phi}) and (\ref{apprIIdel1phi}), and we compute $\varepsilon_1, \delta_1$. After this, it is straightforward to obtain $\varepsilon_2$ and $\delta_2$ using Eq.(~\ref{eps2delta2phi}). Using Eq.(~\ref{apprIIequdNdphi}), we calculate the number of e-folds $N$ by adopting the same technique as we did in the previous subsection (Slow-Roll Approximation I).   Equipped with all the preliminaries, we can finally compute the scalar spectral index ($n_s$), tensor to scalar ratio ($r$), and amplitude of scalar power spectrum ($A_s$) using  Eq.(~\ref{ns_slr}), (\ref{r_slr}), and Eq.(~\ref{As_slr}), respectively. Here, we also take $\xi_1=10$ and $\xi_2 =0.4$; these values of the coupling parameters are chosen so that the constraints on $n_s$ and $r$ do not contradict the current CMB bounds. Whereas $V_0$ can be fixed by matching $A_s=2.09\times10^{-9}$ at the pivot.
 As we have previously mentioned, $n$ and $q$ vary from $1$ to $4$, which results in the $16$ different classes of potential. We have analyzed all the possible combinations. 
In fig.~\ref{rnsslow2}, we present the results on $r-n_s$ where the Slow-Roll I approximation has been utilized. It is evident from fig.~(\ref{rnsslow2}) that almost all the obtained values of $r$ and $n_s$ for a combination of potential parameters $n$ and $q$ are well within the constraints imposed by the {\it Planck'18}. Explicit numerical values of the inflationary observables can be found in Table (\ref{Table2}). A comparison between the two slow-roll approaches for a specific subclass of the PLP model is given in the fig.~(\ref{slow1slow2}).


\begin{widetext}
\begin{center}
\begin{table}[h]
\begin{center}
 \resizebox{\textwidth}{!}{  
\begin{tabular}{|l| l r | l  r | l r | l r |}
\hline
\multicolumn{1}{|c}{ }&\multicolumn{1}{c}{ }&\multicolumn{1}{c}{ }&\multicolumn{1}{c}{ }&\multicolumn{1}{c}{}{$q=1$ }&\multicolumn{1}{c}{}&\multicolumn{1}{c}{}&\multicolumn{1}{c}{}&\multicolumn{1}{c|}{}\\
\hline
\multicolumn{1}{|c|}{ }&\multicolumn{1}{c}{ }{$n=1$ }&\multicolumn{1}{c|}{ }&\multicolumn{1}{c}{ }{$n=2$ }&\multicolumn{1}{c|}{}&\multicolumn{1}{c}{}{$n=3$ }&\multicolumn{1}{c|}{}&\multicolumn{1}{c}{}{$n=4$ }&\multicolumn{1}{c|}{}\\
\hline
\multicolumn{1}{|c|}{$\Delta N$}&\multicolumn{1}{c}{$r$}  &\multicolumn{1}{c|}{ $n_s$ } & \multicolumn{1}{c}{ $r$}&\multicolumn{1}{c|}{{$n_s$}}&\multicolumn{1}{c}{$r$}&\multicolumn{1}{c|}{$n_s$}&\multicolumn{1}{c}{$r$}&\multicolumn{1}{c|}{$n_s$} \\
\hline

 \,\,$50$\,&\,\,0.00729363\,\, &\,\, 0.96551\,\,&\,\,0.0052205\,\,&\,\,0.9611672\,\,&\,\, 0.0049137\,\,&\,\,0.960799\,\,&\,\, 0.00485738\,\,&\,\,0.96068
\\
\
 $60$\,&\,\,0.00535748\,\,&\,\,0.971735\,\,&\,\,0.0037563\,\,&\,\,0.96808\,\,&\,\,0.0034367\,\,&\,\,0.967247\,\,&\,\, 0.00339305\,\,&\,\,0.967135
\\
 \hline
\multicolumn{1}{|c}{ }&\multicolumn{1}{c}{ }&\multicolumn{1}{c}{ }&\multicolumn{1}{c}{ }&\multicolumn{1}{c}{}{$q=2$ }&\multicolumn{1}{c}{}&\multicolumn{1}{c}{}&\multicolumn{1}{c}{}&\multicolumn{1}{c|}{}\\
\hline
\multicolumn{1}{|c|}{ }&\multicolumn{1}{c}{ }{$n=1$ }&\multicolumn{1}{c|}{ }&\multicolumn{1}{c}{ }{$n=2$ }&\multicolumn{1}{c|}{}&\multicolumn{1}{c}{}{$n=3$ }&\multicolumn{1}{c|}{}&\multicolumn{1}{c}{}{$n=4$ }&\multicolumn{1}{c|}{}\\
\hline
\multicolumn{1}{|c|}{$\Delta N$}&\multicolumn{1}{c}{$r$}  &\multicolumn{1}{c|}{ $n_s$ } & \multicolumn{1}{c}{ $r$}&\multicolumn{1}{c|}{{$n_s$}}&\multicolumn{1}{c}{$r$}&\multicolumn{1}{c|}{$n_s$}&\multicolumn{1}{c}{$r$}&\multicolumn{1}{c|}{$n_s$} \\
\hline

 \,\,$50$\,&\,\,0.0103358\,\, &\,\, 0.968584\,\,&\,\,0.00580519\,\,&\,\,0.962561\,\,&\,\, 0.00497474\,\,&\,\,0.960935\,\,&\,\, 0.00486187\,\,&\,\,0.960702
\\
\
 $60$\,&\,\,0.0078487\,\,&\,\,0.974514\,\,&\,\,0.00413578\,\,&\,\,0.968917\,\,&\,\,0.00348436\,\,&\,\,0.967372\,\,&\,\, 0.00339693\,\,&\,\,0.967153
\\

 \hline
 \multicolumn{1}{|c}{ }&\multicolumn{1}{c}{ }&\multicolumn{1}{c}{ }&\multicolumn{1}{c}{ }&\multicolumn{1}{c}{}{$q=3$ }&\multicolumn{1}{c}{}&\multicolumn{1}{c}{}&\multicolumn{1}{c}{}&\multicolumn{1}{c|}{}\\
\hline
\multicolumn{1}{|c|}{ }&\multicolumn{1}{c}{ }{$n=1$ }&\multicolumn{1}{c|}{ }&\multicolumn{1}{c}{ }{$n=2$ }&\multicolumn{1}{c|}{}&\multicolumn{1}{c}{}{$n=3$ }&\multicolumn{1}{c|}{}&\multicolumn{1}{c}{}{$n=4$ }&\multicolumn{1}{c|}{}\\
\hline
\multicolumn{1}{|c|}{$\Delta N$}&\multicolumn{1}{c}{$r$}  &\multicolumn{1}{c|}{ $n_s$ } & \multicolumn{1}{c}{ $r$}&\multicolumn{1}{c|}{{$n_s$}}&\multicolumn{1}{c}{$r$}&\multicolumn{1}{c|}{$n_s$}&\multicolumn{1}{c}{$r$}&\multicolumn{1}{c|}{$n_s$} \\
\hline

 \,\,$50$\,&\,\,0.0138884\,\, &\,\, 0.970306\,\,&\,\,0.00630019\,\,&\,\,0.963339\,\,&\,\, 0.00503582\,\,&\,\,0.961067\,\,&\,\, 0.00486645\,\,&\,\,0.960723
\\
\
 $60$\,&\,\,0.0107645\,\,&\,\,0.975965\,\,&\,\,0.0045256\,\,&\,\,0.969642\,\,&\,\,0.00353203\,\,&\,\,0.967494\,\,&\,\, 0.00340085\,\,&\,\,0.967171
\\
\hline
\multicolumn{1}{|c}{ }&\multicolumn{1}{c}{ }&\multicolumn{1}{c}{ }&\multicolumn{1}{c}{ }&\multicolumn{1}{c}{}{$q=4$ }&\multicolumn{1}{c}{}&\multicolumn{1}{c}{}&\multicolumn{1}{c}{}&\multicolumn{1}{c|}{}\\
\hline
\multicolumn{1}{|c|}{ }&\multicolumn{1}{c}{ }{$n=1$ }&\multicolumn{1}{c|}{ }&\multicolumn{1}{c}{ }{$n=2$ }&\multicolumn{1}{c|}{}&\multicolumn{1}{c}{}{$n=3$ }&\multicolumn{1}{c|}{}&\multicolumn{1}{c}{}{$n=4$ }&\multicolumn{1}{c|}{}\\
\hline
\multicolumn{1}{|c|}{$\Delta N$}&\multicolumn{1}{c}{$r$}  &\multicolumn{1}{c|}{ $n_s$ } & \multicolumn{1}{c}{ $r$}&\multicolumn{1}{c|}{{$n_s$}}&\multicolumn{1}{c}{$r$}&\multicolumn{1}{c|}{$n_s$}&\multicolumn{1}{c}{$r$}&\multicolumn{1}{c|}{$n_s$} \\
\hline

 \,\,$50$\,&\,\,0.0178204\,\, &\,\, 0.971075\,\,&\,\,0.00680496\,\,&\,\,0.964019\,\,&\,\, 0.0050969\,\,&\,\,0.961195\,\,&\,\, 0.00487111\,\,&\,\,0.960743
\\
\
 $60$\,&\,\,0.0139747\,\,&\,\,0.976535\,\,&\,\,0.00492388\,\,&\,\,0.970269\,\,&\,\,0.003579\,\,&\,\,0.967613\,\,&\,\, 0.00340481\,\,&\,\,0.967189
\\
\hline
\end{tabular}}
\end{center}
\caption{Table for $r-n_{s}$ for different values of potential parameters in Slow-Roll II}
\label{Table2}
\end{table}
\end{center}
\end{widetext}

\section{Reheating Parameters}
\label{sec:reh}
After the end of the inflationary epoch, the universe goes into a super-cold mode. In a standard cold inflationary scenario, the only degree of freedom that can survive after the inflation is the inflation field itself. However, to repopulate the universe, an era of reheating is required\cite{PhysRevLett.48.1437,Kofman:1994rk,Martin:2014nya,Martin:2014nya,Mishra:2021wkm,Gialamas:2019nly,Gialamas:2024jeb}, during which the inflaton field transfers its energy to other degrees of freedom, reviving the universe from a super-cooled state to a hot thermal bath of relativistic particles. The concept of reheating was first proposed in \cite{Kofman:1994rk}. In the standard cold inflationary scenario, various methods of reheating have been suggested in the literature. One such method is perturbative decay, where the inflaton field reaches the bottom of its potential and begins to decay into other elementary particles\cite{Kofman:1997yn,Shtanov:1994ce,Bassett:2005xm,Rehagen:2015zma}. The resulting particles interact and reach equilibrium at a temperature known as the reheating temperature ($T_{re}$). More sophisticated methods include parametric resonance, which is a non-perturbative method, tachyonic instability\cite{Felder:2001kt}, and preheating\cite{Lozanov:2019jxc,Kofman:1997pt}. The initial stage of reheating is often attributed to the preheating phase, which is more efficient than perturbative reheating because it involves exponential decay, generating a large number of particles.

However, it has been shown that the reheating era can be studied without delving into its actual dynamics\cite{Cook:2015vqa}. Since there is no direct observational bound on the reheating temperature, analyzing this era indirectly can be extremely useful. This approach allows us to estimate the thermalization temperature using inflationary observables and can also serve as a new way to constrain different inflationary models. In addition to the reheating temperature ($T_{re}$) and the equation of state parameter ($\omega_{re}$), another crucial quantity is the duration of reheating ($N_{re}$). $N_{re}$ quantifies the expansion of the universe from the end of inflation to the end of the reheating era\cite{Cook:2015vqa,Khan:2022odn,Gangopadhyay:2022vgh,Adhikari:2019uaw}.

Assuming $w_{re}$ to be constant during the reheating epoch, the energy density of the universe at the end of inflation can be expressed in terms of the scale factor through  $\rho \propto a^{-3(1+w)}$, formulated as follows:
\begin{equation}
    \frac{\rho_{end}}{\rho_{re}} = \left(\frac{a_{end}}{a_{re}} \right)^{-3(1+w_{re})},
    \label{re1}
\end{equation}
The subscript $end$ signifies the end of inflation, while $re$ stands for the end of the reheating period. Using the end of inflation condition($\varepsilon=1$). 

Here crucial point is to calculate the $\rho_{end}$ for the EGB gravity. In standard cold inflationary scenarios, it is straightforward to calculate $\rho_{end}$ by employing $\omega = \frac{P}{\rho}= -\frac{1}{3}$. However, in EGB gravity, the end of inflation might not occur at $\omega= -\frac{1}{3}$, and it can be slightly different. From Eq. \ref{Equ00} (keeping  $U_0=1/2$)
\begin{eqnarray}
    3 H^2 = \frac{\psi^2}{2}+ V+12  \xi_{,\phi}\psi H^3
\end{eqnarray}
Comparing this with the standard cold inflationary scenario where $3 H^2 = \rho_{\phi} =\frac{\psi^2}{2}+ V$ we can compute the extra contribution to the energy density due to the modified gravity($\rho_{egb}=  \xi_{,\phi}\psi H^3$).
\begin{equation}
   3 H^2 = \rho_{\phi} +\rho_{egb} 
\end{equation}
Similarly from Eq. \ref{Equ11} we can write
\begin{equation}
   2 \dot{H}=-\left( \psi^2- 4 H^2 \xi_{,\phi\phi} \psi^2 - 4 H^2 \xi_{,\phi} \dot{\psi} + 4H^3 \xi_{,\phi} \psi- 8\xi_{,\phi} \psi H \dot{H}  \right)
\end{equation}
Now comparing this with $2\dot{H} = -(P_{\phi}+\rho_{\phi})$ we can compute the extra contribution to the pressure due to the presence of the EGB gravity. We write
\begin{equation}
  2 \dot{H} = - \Big(  \left(P_{\phi}+ P_{egb} \right)+ \left( \rho_{\phi}+ \rho_{egb} \right)  \big)
\end{equation}
where, $P_{egb} = -4 H^2 \xi,_{\phi \phi}\psi^2- 4 H^3 \xi,_{\phi} \dot{\psi}- \frac{8}{3} H^3 \xi,_{\phi} \psi - 8 \xi,_{\phi} \psi H \dot{ H}$, now we calculate the $\rho_{\phi}$ in terms of the inflationary potential from $\omega_{\phi} = \frac{P_{\phi}}{\rho_{\phi}}$ at the end of inflation.
\begin{equation}
    \rho_{\phi} =V \left( \frac{2}{1-\omega_{\phi}} \right)
\end{equation}
Also, we are writing $\frac{\rho_{egb}}{\rho_{\phi}}= \alpha$, this allows us to write the total energy density at the end of inflation as:
\begin{equation}
  \rho_{\rm total} = \rho_{\phi}+ \rho_{egb} =\rho_{\phi} (1+ \alpha) = \beta V, ~~~~~~~ {\rm where }~~~~  \beta= \left( \frac{2}{1-\omega_{\phi}}\right) \left( 1+ \alpha  \right)    
\end{equation}
 
Substituting it in eq. (\ref{re1}), we find
\begin{equation}
    N_{re} = \frac{1}{3(1+w_{re})} \ln \left(\frac{\rho_{end}}{\rho_{re}} \right)= \frac{1}{3(1+w_{re})} \ln \left(\textcolor{blue}{\beta}\frac{ V_{end}}{\rho_{re}} \right),
    \label{re2}
\end{equation}
Also, we know:
\begin{equation}
\rho_{re} = \frac{\pi^2}{30} g_{re} T_{re}^4.
\label{re3}
\end{equation}
Where $g_{re}$ denotes the number of relativistic species at the end of reheating.
Using (\ref{re2}) and (\ref{re3}) and following  \cite{Cook:2015vqa, Cai:2015soa, Gong:2015qha}, we can express $T_{re}$ and $N_{re}$ as :
\begin{equation}
N_{re} = \frac{1}{3(1+w_{re})} \ln \left(\frac{30~ {\beta}~ V_{end}}{\pi^2 g_{re} T_{re}^4 } \right)
\label{re4}
\end{equation}
Admitting that entropy is conserved from the reheating to today, one may write  
\begin{equation}
T_{re}= T_0 \left(\frac{a_0}{a_{re}} \right) \left(\frac{43}{11 g_{re}} \right)^{\frac{1}{3}}=T_0 \left(\frac{a_0}{a_{eq}} \right) e^{N_{RD}} \left(\frac{43}{11 g_{re}} \right)^{\frac{1}{3}},
\label{re5}
\end{equation}
with $N_{RD}$ being the number of e-folds during radiation era, since $e^{-N_{RD}}\equiv a_{re}/a_{eq}$. The ratio $a_{0}/a_{eq}$ can be expressed as 
\begin{equation}
\frac{a_0}{a_{eq}} = \frac{a_0 H_{k}}{k} e^{-N_{k}} e^{- N_{re}} e^{- N_{RD}}\
\label{re6}
\end{equation}
We know for the modes at the horizon exit we can write, $k_{}=a_{k} H_{k}$ and using the Eq.~(\ref{re4}), (\ref{re5}) and (\ref{re6}), assuming $w_{re} \neq \frac{1}{3}$ and $g_{re} \approx 226$ (degrees of freedom in a supersymmetric model), we can derive the expression for $N_{re}$
\begin{equation}
N_{re}= \frac{4}{ (1-3w_{re} )}   \left[{61.488}  - \ln \left(\frac{ V_{end}^{\frac{1}{4}}}{ H_{k} } \right)  - N_{k}   \right]
\label{re7}
\end{equation}
Using Planck's pivot  ($k$) of order $0.05 \; \mbox{Mpc}^{-1}$ we find $T_{re}$: 
\begin{equation}
T_{re}= \left[ \left(\frac{43}{11 g_{re}} \right)^{\frac{1}{3}}    \frac{a_0 T_0}{k_{}} H_{k} e^{- N_{k}} \left[\frac{30~ {\beta}~ V_{end}}{\pi^2 g_{re}} \right]^{- \frac{1}{3(1 + w_{re})}}  \right]^{\frac{3(1+ w_{re})}{3 w_{re} -1}}.
\label{re8}
\end{equation}
    
It is noteworthy to mention that the expressions of $N_{re}$ and $T_{re}$ obtained here are general expressions. Which are valid for the EGB gravity, also all the information of EGB gravity is encoded in the $\beta$, $n_s, N_k$, and $H_k$ through the coupling $\xi (\phi)$.  Evolution and the converging point of $N_{re}$ and $T_{re}$ are different for both the slow roll approximation and the exact solution. This is because the evolution of $n_s, N_k$, and $H_k$ is slightly different in approximations and numerical solutions. However, the order of the quantity $\beta$ is always around $\mathcal{O}(1)$, which was also the case in the standard cold inflation scenario. Thus, the inclusion of $\beta$ has a negligible effect on the behavior of $N_{re}$ and $T_{re}$. Also, the first term in the square bracket in Eq. \ref{re7} has a contribution from $\beta$, but this contribution is also minimal. From Eq.(\ref{re7}) and (\ref{re8}) it is evident that to evaluate $N_{re}$ and $T_{re}$, first, we require to calculate the $H_{k}$, $N_{k}$ (e-folds during the inflation) and $V_{end}$ (value of the potential at the end of inflation). Utilizing the tensor-to-scalar ratio with the scalar power spectrum,m we can write
\begin{equation}
{H_k}=\sqrt{\frac{1}{2} \pi ^2 A_{s}  r}.
\label{Hk}
\end{equation}
Maintaining $A_s(k_0)= 2.0989\times10^{-9}$, it is evident from Eq.(\ref{re7}) and (\ref{re8}) that both equations depend on $H_k$.

From Eq.(\ref{Hk}), it's clear that $H_k$ is a function of the tensor-to-scalar ratio ($r$). As we know from the CMB observation, tensor to scalar ratio has only an upper bound. To precisely analyze the dynamics of reheating, we need to express $H_k$ in terms of the spectral index ($n_s$). This can be done from the relation between $r,n_s$ and $N_k$\cite{Adhikari:2019uaw,Khan:2022odn}. As mentioned in \cite{Cook:2015vqa}, varying $A_s$ with $n_s$ has negligible effect on the reheating, so we keep $A_s(k_0)= 2.0989\times10^{-9}$. 
\begin{figure}[!htb]
\centering
\includegraphics[width=8cm,height=8cm]{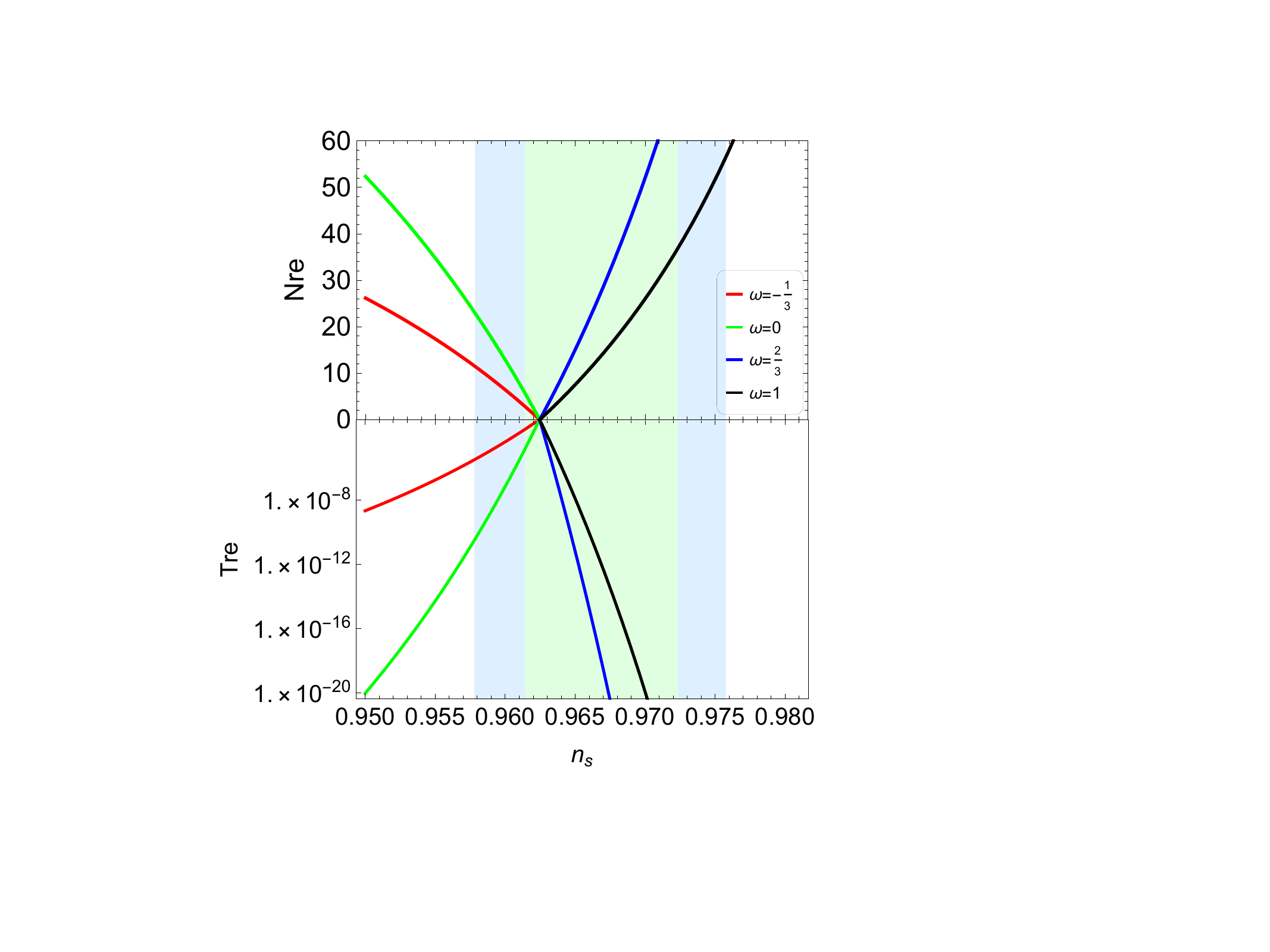}
\includegraphics[width=8cm,height=8cm]{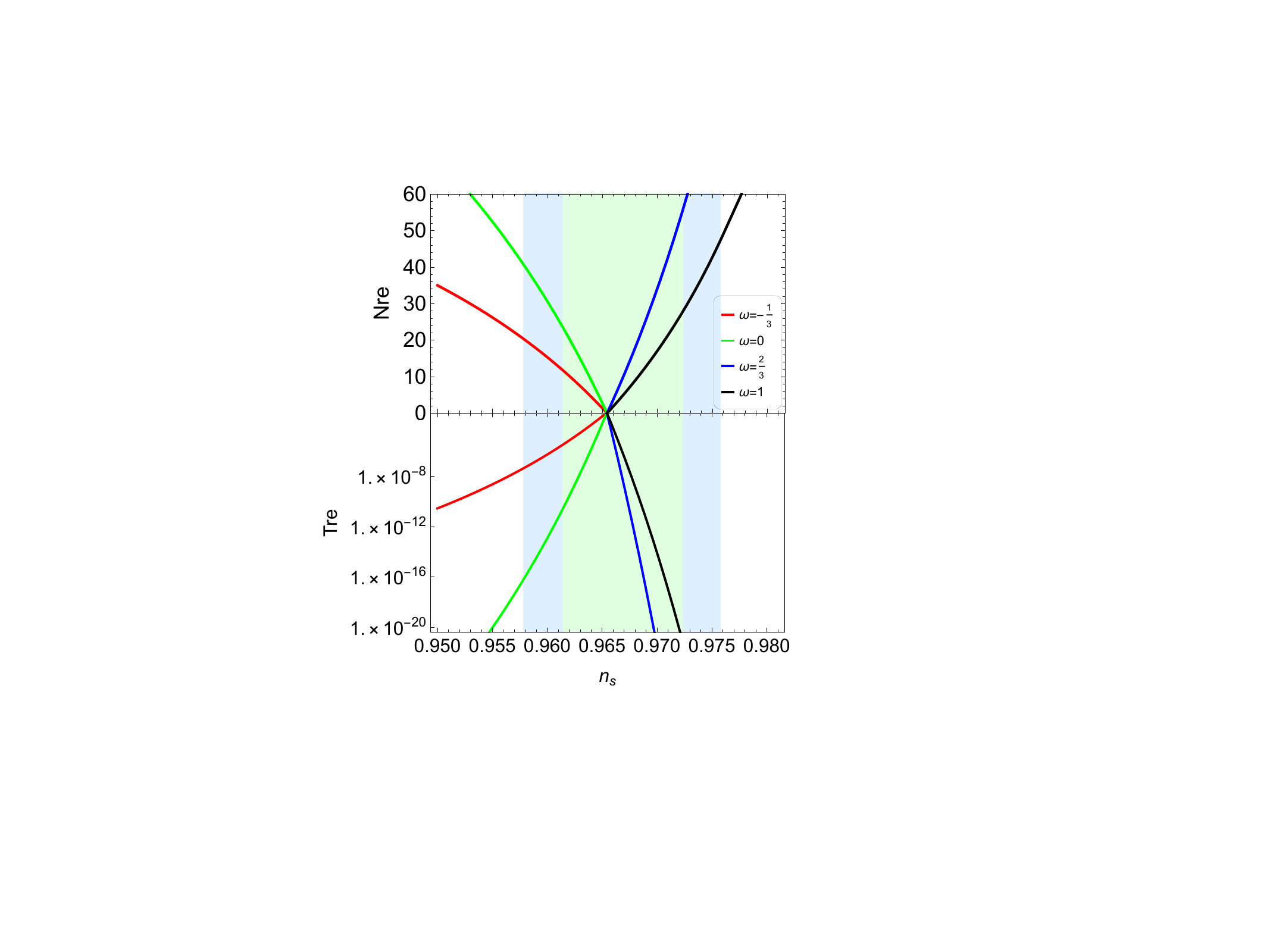}
\caption{Plots illustrating $N_{re}$ and $T_{re}$ for different choices of $\omega_{re}$ are presented. 
\textbf{Left Panel:} The results for slow-roll approximation $\text{I}$ are represented. 
\textbf{Right Panel:}  The results for slow-roll approximation $\text{II}$ are represented. The light blue shaded area is for the $2-\sigma$ bounds, while the light green shaded area represents the $1-\sigma$ bounds on $n_s$ from Planck'18 \cite{Planck:2018jri}. Finally, the self-explanatory color coding for $w_{re}$ is given in the inset of the plots.}
\label{reheating_plots}
\end{figure}

\begin{figure}
    \centering
    \includegraphics[width=0.5\linewidth]{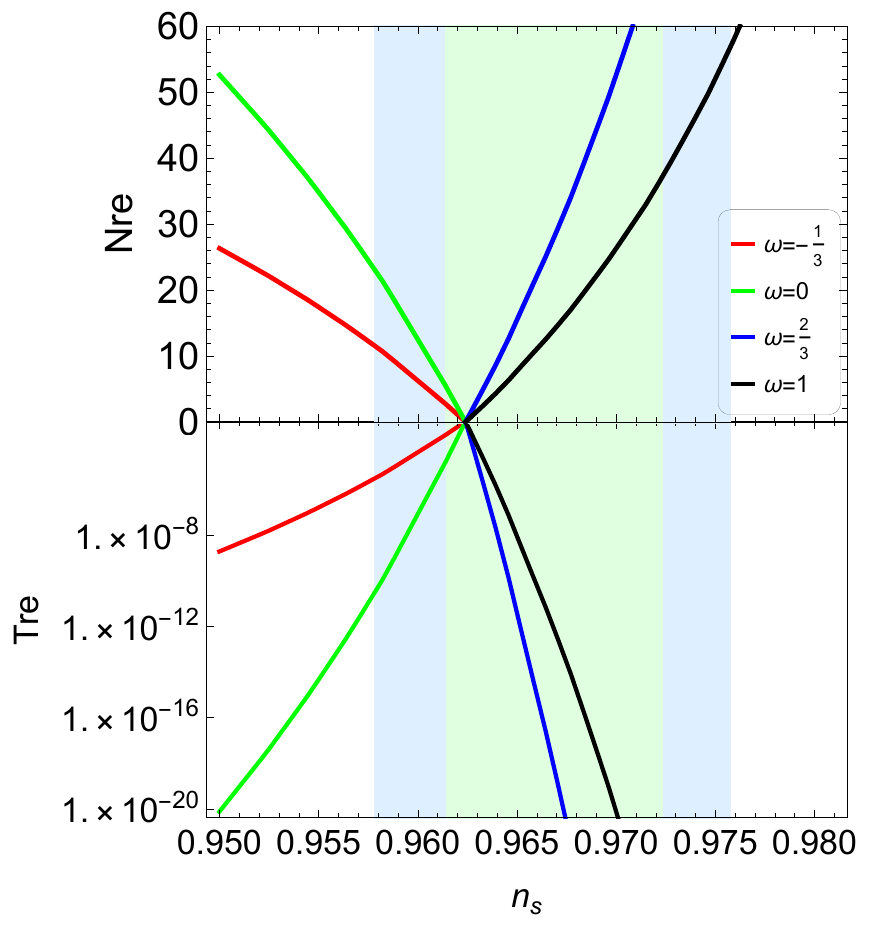}
    \caption{Plots illustrating $N_{re}$ and $T_{re}$ for different choices of $\omega_{re}$  by solving system (~\ref{DynSYSN}) numerically for $n=2$ and $q=1$. Plot legends are self-explanatory.}
    \label{tre_nre_num}
\end{figure}

 The variation of $N_{re}$ and $T_{re}$ for the PLP model with $n=2$ and $q=1$ for various values of $w_{re}$ is shown in Fig.~(\ref{reheating_plots}). In Fig. \ref{tre_nre_num} we present the reheating analysis for the exact solution obtained by numerically solving Eq. (~\ref{DynSYSN}). The convergence points on the $T_{re}$ and $N_{re}$ plots represent instantaneous reheating, with $N_{re}=0$. One can, of course, do this exercise for all 16 combinations for this class of inflationary model. But the essence of the findings can be encapsulated in one of these figures, such as fig.~(\ref{reheating_plots}) and (\ref{tre_nre_num}).

\section{Conclusions and Discussions}
\label{discussions}
\setcounter{equation}{0}
\setcounter{figure}{0}
\setcounter{table}{0}
\setcounter{footnote}{1}

With future observations like {\it CMB-S4} \cite{CMB-S4:2016ple} and {\it COrE} \cite{CORE:2016ymi}, which have promising prospects to measure the spectral tilt very precisely ($\Delta n_s \sim 0.002$), and with future possibilities to constrain the primordial tensor modes, a revisit to theoretically well-motivated models is in order. Not only that, the conceptual development in the modified gravity sector in the last few years has opened a window to study several interesting features. In the case of EGB gravity, new developments such as the new slow roll approximation methods proposed in \cite{Pozdeeva:2024ihc}, beg for a more detailed analysis. In this work, we had a threefold objective:
\begin{itemize}
    \item Rescue a theoretically well-motivated model using a non-standard gravitational background of EGB gravity, which we show can be done successfully.
    \item Implement the newly proposed idea of slow-roll approximation in the case of the PLP model and check the difference in predictivity depending on the approximation method.
    \item Finally, in the light of a more realistic case of generalized reheating, how the inflationary predictivity enlightens us on a model in the EGB background. 
\end{itemize}
Thus, using this prescription, we have summarised a path to study inflationary dynamics in the EGB gravity. We have shown that the class of PLP inflationary dynamics in the EGB background can successfully yield observationally astute results, even keeping the generalized reheating in mind. In our analysis, we found field values at the start of inflation in Slow Roll Approximation-I are close to the field values obtained by numerically solving Eq.~\ref{DynSYSN} (see Table \ref{phi_tabslow1},\ref{phi_tabslow2} and \ref{phi_tab_numeric}). Thus, we conclude that in our study, Slow Roll Approximation-I is more accurate than the Slow Roll Approximation-II 

A few more studies that we plan to conduct in the EGB inflationary framework will be to see how the dynamics of Warm inflation get incorporated \cite{Bastero-Gil:2016qru,Bastero-Gil:2017wwl}. Production of Primordial Black Holes and Gravitational Waves in case of warm inflationary dynamics \cite{Gangopadhyay:2020bxn, Basak:2021cgk, Correa:2023whf,Correa:2022ngq} in the EGB background, utilizing this new prescription could be a really interesting path to follow. The authors would like to address these questions in future studies.

\section*{Acknowledgments} 
The authors would like to thank Mostafizur Rahman, Arun Kumar, and Abolhassan Mohammadi for useful discussions. The work of MRG is supported by the Department of Science and Technology(DST), Government of
India under the Grant Agreement number IF18-PH-228 (INSPIRE Faculty Award) and by the Science and Engineering
Research Board(SERB), DST, Government of India, under the Grant Agreement number CRG/2022/004120(CoreResearch Grant). MRG would like to thank IUCAA, Pune, India, for their hospitality during the visit as an associate when the project was initiated.
MRG would also like to thank OMEG Institute, Soongsil University, Republic of Korea, for their hospitality, where the paper's final part was completed.

\bibliographystyle{apsrev4-1}
\bibliography{PLP_EGB_R1.bib}


\newpage
\appendix

\section{Explicit Analytic Forms of Slow-Roll Parameters}
\label{App}
\subsection{\underline{New Slow Roll Approximation I}}
\label{App1}
The slow-roll parameters in the case of slow-roll approximation I are given below:
\begin{equation*}
\delta_1=-\frac{\text{$\xi_1$} \text{$\xi_2$} \text{sech}^2(\text{$\xi_2$} \phi ) \left(\frac{\phi ^n}{m^n+\phi ^n}\right)^{2 q} \left(3 n q m^n \phi ^{-n-1} \left(\frac{\phi ^n}{m^n+\phi ^n}\right)^{1-q}+4 \text{$\xi_1$} \text{$\xi_2$} \text{sech}^2(\text{$\xi_2$} \phi )\right)}{6 \left(\text{$\xi_1$}^2 \text{$\xi_2$}^2 \text{sech}^4(\text{$\xi_2$} \phi ) \left(\frac{\phi ^n}{m^n+\phi ^n}\right)^{2 q}+\frac{3}{8}\right)}
\end{equation*}
\begin{eqnarray*}
    \delta_2&=&  \Bigg[9 \cosh ^4\left(\xi _2 \phi \right) \Bigg(-3 n q m^n \cosh ^4\left(\xi _2 \phi \right) \left(m^n (n q-1)-(n+1) \phi ^n\right) n q \phi  +m^n \left(m^n+\phi ^n\right) \cosh ^2\left(\xi _2 \phi \right)\nonumber \\ && \xi _2 \left(3 \sinh \left(2 \xi _2 \phi \right)-8 \xi _1 \left(\frac{\phi ^n}{m^n+\phi ^n}\right)^q\right)+8 \xi _1 \left(\frac{\phi ^n}{m^n+\phi ^n}\right)^q \bigg( \phi ^2 \left(m^n+\phi ^n\right)^2 \sinh \left(2 \xi _2 \phi \right)+n q m^n \left(\frac{\phi ^n}{m^n+\phi ^n}\right)^q \nonumber \\&& \xi _1 \left(m^n (n q+1)+(n+1) \phi ^n\right) \bigg)\xi_2^2 -16 n \xi _1^2 \xi _2^3 q \phi  m^n \left(m^n+\phi ^n\right) \tanh \left(\xi _2 \phi \right) \left(\frac{\phi ^n}{m^n+\phi ^n}\right)^{2 q}\Bigg)\Bigg]\Bigg/ \Bigg[ \phi  \left(m^n+\phi ^n\right) \times \nonumber \\&& \Bigg( 8 \xi _1^2 \xi _2^2 \left(\frac{\phi ^n}{m^n+\phi ^n}\right)^{2 q}+3 \cosh ^4\left(\xi _2 \phi \right)  \Bigg) \Bigg( 9 \phi  \left(m^n+\phi ^n\right) \cosh ^4\left(\xi _2 \phi \right)-12 n \xi _1 \xi _2 q m^n \cosh ^2\left(\xi _2 \phi \right) \left(\frac{\phi ^n}{m^n+\phi ^n}\right)^q \nonumber \\&&+8 \xi _1^2 \xi _2^2 \phi  \left(m^n+\phi ^n\right) \left(\frac{\phi ^n}{m^n+\phi ^n}\right)^{2 q}  \Bigg) \Bigg]
\end{eqnarray*}
\begin{widetext}
\begin{eqnarray}
     \varepsilon_1&=&  \Bigg[ 27 \left(\frac{\phi ^n}{m^n+\phi ^n}\right)^{-2 q} \left(\xi _1^2 \xi _2^2 \text{sech}^4\left(\xi _2 \phi \right) \left(\frac{\phi ^n}{m^n+\phi ^n}\right)^{2 q}+\frac{3}{8}\right) \Bigg(\xi _1 \xi _2 V_0 \text{sech}^2\left(\xi _2 \phi \right) \left(\frac{\phi ^n}{m^n+\phi ^n}\right)^{2 q}\nonumber\\&&+\frac{3}{4} n q V_0 m^n \phi ^{-n-1} \left(\frac{\phi ^n}{m^n+\phi ^n}\right)^{q+1}  \Bigg) \Bigg(  \frac{q V_0 \left(\frac{n \phi ^{n-1}}{m^n+\phi ^n}-\frac{n \phi ^{2 n-1}}{\left(m^n+\phi ^n\right)^2}\right) \left(\frac{\phi ^n}{m^n+\phi ^n}\right)^{q-1}}{9 \left(\xi _1^2 \xi _2^2 \text{sech}^4\left(\xi _2 \phi \right) \left(\frac{\phi ^n}{m^n+\phi ^n}\right)^{2 q}+\frac{3}{8}\right)} \nonumber \\&& \times \left(\xi _1^2 \xi _2^2 \text{sech}^4\left(\xi _2 \phi \right) \left(\frac{\phi ^n}{m^n+\phi ^n}\right)^{2 q}-\frac{1}{2} 3 n \xi _1 \xi _2 q m^n \phi ^{-n-1} \text{sech}^2\left(\xi _2 \phi \right) \left(\frac{\phi ^n}{m^n+\phi ^n}\right)^{q+1}+\frac{9}{8} \right) \nonumber \\&& - \frac{V_0 \left(\frac{\phi ^n}{m^n+\phi ^n}\right)^q}{9 \left(\xi _1^2 \xi _2^2 \text{sech}^4\left(\xi _2 \phi \right) \left(\frac{\phi ^n}{m^n+\phi ^n}\right)^{2 q}+\frac{3}{8}\right){}^2} \Bigg( \xi _1^2 \xi _2^2 \text{sech}^4\left(\xi _2 \phi \right) \left(\frac{\phi ^n}{m^n+\phi ^n}\right)^{2 q}-\frac{1}{2} 3 n \xi _1 \xi _2 q m^n \phi ^{-n-1} \text{sech}^2\left(\xi _2 \phi \right) \left(\frac{\phi ^n}{m^n+\phi ^n}\right)^{q+1} \nonumber \\ && +\frac{9}{8}  \Bigg) \left(2 \xi _1^2 \xi _2^2 q \left(\frac{n \phi ^{n-1}}{m^n+\phi ^n}-\frac{n \phi ^{2 n-1}}{\left(m^n+\phi ^n\right)^2}\right) \text{sech}^4\left(\xi _2 \phi \right) \left(\frac{\phi ^n}{m^n+\phi ^n}\right)^{2 q-1} -4 \xi _1^2 \xi _2^3 \tanh \left(\xi _2 \phi \right) \text{sech}^4\left(\xi _2 \phi \right) \left(\frac{\phi ^n}{m^n+\phi ^n}\right)^{2 q}\right)\nonumber \\&& + \frac{\left(\frac{\phi ^n}{m^n+\phi ^n}\right)^q}{9 \left(\xi _1^2 \xi _2^2 \text{sech}^4\left(\xi _2 \phi \right) \left(\frac{\phi ^n}{m^n+\phi ^n}\right)^{2 q}+\frac{3}{8}\right)} V_0        \Bigg( \frac{1}{2} (-3) (-n-1) n \xi _1 \xi _2 q m^n \phi ^{-n-2} \text{sech}^2\left(\xi _2 \phi \right) \left(\frac{\phi ^n}{m^n+\phi ^n}\right)^{q+1}- \nonumber \\&&  \frac{3}{2} n \xi _1 \xi _2 q (q+1) m^n \phi ^{-n-1} \left(\frac{n \phi ^{n-1}}{m^n+\phi ^n}-\frac{n \phi ^{2 n-1}}{\left(m^n+\phi ^n\right)^2}\right) \text{sech}^2\left(\xi _2 \phi \right) \left(\frac{\phi ^n}{m^n+\phi ^n}\right)^q+2 q \left(\frac{\phi ^n}{m^n+\phi ^n}\right)^{2 q-1} \nonumber \\&& \xi _1^2 \xi _2^2 \left(\frac{n \phi ^{n-1}}{m^n+\phi ^n}-\frac{n \phi ^{2 n-1}}{\left(m^n+\phi ^n\right)^2}\right) \text{sech}^4\left(\xi _2 \phi \right)+3 n \xi _1 \xi _2^2 q m^n \phi ^{-n-1} \tanh \left(\xi _2 \phi \right) \text{sech}^2\left(\xi _2 \phi \right) \left(\frac{\phi ^n}{m^n+\phi ^n}\right)^{q+1} \nonumber \\&& -4 \xi _1^2 \xi _2^3 \tanh \left(\xi _2 \phi \right) \text{sech}^4\left(\xi _2 \phi \right) \left(\frac{\phi ^n}{m^n+\phi ^n}\right)^{2 q} \Bigg) \Bigg)\Bigg]\Bigg/ \nonumber\\ && 4 V_0^2 \left(\xi _1^2 \xi _2^2 \text{sech}^4\left(\xi _2 \phi \right) \left(\frac{\phi ^n}{m^n+\phi ^n}\right)^{2 q}-\frac{1}{2} 3 n \xi _1 \xi _2 q m^n \phi ^{-n-1} \text{sech}^2\left(\xi _2 \phi \right) \left(\frac{\phi ^n}{m^n+\phi ^n}\right)^{q+1}+\frac{9}{8}\right){}^2 
\end{eqnarray}
\end{widetext}   

\begin{eqnarray}
    \varepsilon_2&=&\bigg[3 \left(\frac{\phi ^n}{m^n+\phi ^n}\right)^q \left(3 n q m^n \phi ^{-n-1} \left(\frac{\phi ^n}{m^n+\phi ^n}\right)^{1-q}+4 \xi _1 \xi _2 \text{sech}^2\left(\xi _2 \phi \right)\right)\bigg(-\bigg(\bigg(4 \xi _1 \xi _2 \phi  \left(m^n+\phi ^n\right) \text{sech}^2\left(\xi _2 \phi \right) \left(\frac{\phi ^n}{m^n+\phi ^n}\right)^q\nonumber\\&&+3 n q m^n\bigg)\bigg(8 \xi _1^2 \xi _2^2 \phi  \left(m^n+\phi ^n\right) \text{sech}^4\left(\xi _2 \phi \right) \left(\frac{\phi ^n}{m^n+\phi ^n}\right)^{2 q}-12 n \xi _1 \xi _2 q m^n \text{sech}^2\left(\xi _2 \phi \right) \left(\frac{\phi ^n}{m^n+\phi ^n}\right)^q+9 \phi  \left(m^n+\phi ^n\right)\bigg)\nonumber\\&& \times \bigg(81 n q \phi  m^n \left(m^n+\phi ^n\right) \left(m^n (n q-1)-(n+1) \phi ^n\right)+512m^n n q\phi \left(\frac{\phi ^n}{m^n+\phi ^n}\right)^{6 q}(m^n+\phi ^n)\left(m^n (n q-1)-(n+1) \phi ^n\right)\nonumber\\&&\xi _1^6 \xi _2^6 \text{sech}^{12}\left(\xi _2 \phi \right)-54 n q m^n \left(\frac{\phi ^n}{m^n+\phi ^n}\right)^q\xi _1 \xi _2 \text{sech}^4\left(\xi _2 \phi \right)\bigg(2\bigg(m^{2 n} \left(4 n^2 q^2-6 n q+2\right)+(n+1) \left(-m^n\right) (6 n q+n-4) \phi ^n\nonumber\\&&+\left(n^2+3 n+2\right) \phi ^{2 n}\bigg)\cosh ^2\left(\xi _2 \phi \right)-4 \xi _2 \phi  \left(m^n+\phi ^n\right) \sinh \left(2 \xi _2 \phi \right) \left(m^n (2 n q-1)-(n+1) \phi ^n\right)+4 \xi _2^2 \phi ^2 \left(m^n+\phi ^n\right)^2\nonumber\\&& \left(\cosh \left(2 \xi _2 \phi \right)-2\right)\bigg)-36 \xi _1^2 \xi _2^2 \phi  \left(m^n+\phi ^n\right) \text{sech}^6\left(\xi _2 \phi \right) \left(\frac{\phi ^n}{m^n+\phi ^n}\right)^{2 q}\bigg(16 \xi _2^2 \phi ^2 \left(m^n+\phi ^n\right)^2 \left(2 \cosh \left(2 \xi _2 \phi \right)-3\right)\nonumber\\&&-48 n \xi _2 q \phi  m^n \left(m^n+\phi ^n\right) \sinh \left(2 \xi _2 \phi \right)+6 n q m^n \cosh ^2\left(\xi _2 \phi \right) \left(m^n (3 n q+1)+(n+1) \phi ^n\right)\bigg)+192 \xi _1^4 \xi _2^4 \phi  \left(m^n+\phi ^n\right) \text{sech}^{10}\left(\xi _2 \phi \right)\nonumber\\&& \left(\frac{\phi ^n}{m^n+\phi ^n}\right)^{4 q}\bigg(8 \xi _2^2 \phi ^2 \left(m^n+\phi ^n\right)^2 \left(2 \cosh \left(2 \xi _2 \phi \right)-1\right)-8 n \xi _2 q \phi  m^n \left(m^n+\phi ^n\right) \sinh \left(2 \xi _2 \phi \right)+n q m^n \cosh ^2\left(\xi _2 \phi \right)\nonumber\\&& \left(m^n (5 n q-1)-(n+1) \phi ^n\right)\bigg)+288 n \xi _1^3 \xi _2^3 q m^n \text{sech}^8\left(\xi _2 \phi \right) \left(\frac{\phi ^n}{m^n+\phi ^n}\right)^{3 q}\bigg(2\bigg(m^{2 n} \left(2 n^2 q^2+3 n q-2\right)+(n+1) m^n \nonumber\\&&(3 n q+n-4) \phi ^n-\left(n^2+3 n+2\right) \phi ^{2 n}\bigg)24 \xi _2^2 \phi ^2 \left(m^n+\phi ^n\right)^2 \sinh ^2\left(\xi _2 \phi \right)-12 n \xi _2 q \phi  m^n \left(m^n+\phi ^n\right) \sinh \left(2 \xi _2 \phi \right)+\cosh ^2\left(\xi _2 \phi \right)\bigg)\bigg)\bigg)\nonumber\\&&+2 n q m^n\left(4 \xi _1 \xi _2 \phi  \left(m^n+\phi ^n\right) \text{sech}^2\left(\xi _2 \phi \right) \left(\frac{\phi ^n}{m^n+\phi ^n}\right)^q+3 n q m^n\right)\left(8 \xi _1^2 \xi _2^2 \text{sech}^4\left(\xi _2 \phi \right) \left(\frac{\phi ^n}{m^n+\phi ^n}\right)^{2 q}+3\right)\bigg(8 \xi _1^2 \xi _2^2 \phi  \left(m^n+\phi ^n\right) \nonumber\\&&\text{sech}^4\left(\xi _2 \phi \right) \left(\frac{\phi ^n}{m^n+\phi ^n}\right)^{2 q}-12 n \xi _1 \xi _2 q m^n \text{sech}^2\left(\xi _2 \phi \right) \left(\frac{\phi ^n}{m^n+\phi ^n}\right)^q+9 \phi  \left(m^n+\phi ^n\right)\bigg)\bigg(27 n q \phi  m^n \left(m^n+\phi ^n\right)\nonumber\\&&+64 n \xi _1^4 \xi _2^4 q \phi  m^n \left(m^n+\phi ^n\right) \text{sech}^8\left(\xi _2 \phi \right) \left(\frac{\phi ^n}{m^n+\phi ^n}\right)^{4 q}+192 \xi _1^2 \xi _2^3 \phi ^2 \left(m^n+\phi ^n\right)^2 \tanh \left(\xi _2 \phi \right) \text{sech}^4\left(\xi _2 \phi \right) \left(\frac{\phi ^n}{m^n+\phi ^n}\right)^{2 q}\nonumber\\&&-36 n \xi _1 \xi _2 q m^n \text{sech}^2\left(\xi _2 \phi \right) \left(\frac{\phi ^n}{m^n+\phi ^n}\right)^q \left(-2 \xi _2 \phi  \left(m^n+\phi ^n\right) \tanh \left(\xi _2 \phi \right)+m^n (2 n q-1)-(n+1) \phi ^n\right)-96 n \xi _1^3 \xi _2^3 q m^n \text{sech}^6\left(\xi _2 \phi \right)\nonumber\\&& \left(\frac{\phi ^n}{m^n+\phi ^n}\right)^{3 q} \left(2 \xi _2 \phi  \left(m^n+\phi ^n\right) \tanh \left(\xi _2 \phi \right)-m^n-(n+1) \phi ^n\right)\bigg)+16 \xi _1^2 \xi _2^2 \text{sech}^4\left(\xi _2 \phi \right) \left(\frac{\phi ^n}{m^n+\phi ^n}\right)^{2 q} \bigg(4 \xi _1 \xi _2 \phi  \left(m^n+\phi ^n\right)\nonumber\\&& \text{sech}^2\left(\xi _2 \phi \right) \left(\frac{\phi ^n}{m^n+\phi ^n}\right)^q+3 n q m^n\bigg)\bigg(8 \xi _1^2 \xi _2^2 \phi  \left(m^n+\phi ^n\right) \text{sech}^4\left(\xi _2 \phi \right) \left(\frac{\phi ^n}{m^n+\phi ^n}\right)^{2 q}-12 n \xi _1 \xi _2 q m^n \text{sech}^2\left(\xi _2 \phi \right) \left(\frac{\phi ^n}{m^n+\phi ^n}\right)^q\nonumber\\&&+9 \phi  \left(m^n+\phi ^n\right)\bigg)\left(2 \xi _2 \phi  \left(m^n+\phi ^n\right) \tanh \left(\xi _2 \phi \right)-n q m^n\right)\bigg(64 n \xi _1^4 \xi _2^4 q \phi  m^n \left(m^n+\phi ^n\right) \text{sech}^8\left(\xi _2 \phi \right) \left(\frac{\phi ^n}{m^n+\phi ^n}\right)^{4 q}-96 n \xi _1^3 \xi _2^3 q m^n\nonumber\\&& \text{sech}^6\left(\xi _2 \phi \right) \left(\frac{\phi ^n}{m^n+\phi ^n}\right)^{3 q} \left(2 \xi _2 \phi  \left(m^n+\phi ^n\right) \tanh \left(\xi _2 \phi \right)-m^n-(n+1) \phi ^n\right)-36 n \xi _1 \xi _2 q m^n \text{sech}^2\left(\xi _2 \phi \right) \left(\frac{\phi ^n}{m^n+\phi ^n}\right)^q\nonumber\\&& \left(-2 \xi _2 \phi  \left(m^n+\phi ^n\right) \tanh \left(\xi _2 \phi \right)+m^n (2 n q-1)-(n+1) \phi ^n\right)+192 \xi _1^2 \xi _2^3 \phi ^2 \left(m^n+\phi ^n\right)^2 \tanh \left(\xi _2 \phi \right) \text{sech}^4\left(\xi _2 \phi \right) \left(\frac{\phi ^n}{m^n+\phi ^n}\right)^{2 q}\nonumber\\&&+27 n q \phi  m^n \left(m^n+\phi ^n\right)\bigg)+8 \xi _1 \xi _2 \text{sech}^2\left(\xi _2 \phi \right) \left(\frac{\phi ^n}{m^n+\phi ^n}\right)^q\bigg(4 \xi _1 \xi _2 \phi  \left(m^n+\phi ^n\right) \text{sech}^2\left(\xi _2 \phi \right) \left(\frac{\phi ^n}{m^n+\phi ^n}\right)^q+3 n q m^n\bigg)\nonumber\\&&\left(8 \xi _1^2 \xi _2^2 \text{sech}^4\left(\xi _2 \phi \right) \left(\frac{\phi ^n}{m^n+\phi ^n}\right)^{2 q}+3\right)\bigg(4 \xi _1 \xi _2 \phi  \left(m^n+\phi ^n\right) \text{sech}^2\left(\xi _2 \phi \right) \left(\frac{\phi ^n}{m^n+\phi ^n}\right)^q \left(n q m^n-2 \xi _2 \phi  \left(m^n+\phi ^n\right) \tanh \left(\xi _2 \phi \right)\right)\nonumber\\&&-3 n q m^n \left(-2 \xi _2 \phi  \left(m^n+\phi ^n\right) \tanh \left(\xi _2 \phi \right)+m^n (n q-1)-(n+1) \phi ^n\right)\bigg)\bigg(64 n \xi _1^4 \xi _2^4 q \phi  m^n \left(m^n+\phi ^n\right) \text{sech}^8\left(\xi _2 \phi \right) \left(\frac{\phi ^n}{m^n+\phi ^n}\right)^{4 q}\nonumber\\&&+27 n q \phi  m^n \left(m^n+\phi ^n\right)+192 \xi _1^2 \xi _2^3 \phi ^2 \left(m^n+\phi ^n\right)^2 \tanh \left(\xi _2 \phi \right) \text{sech}^4\left(\xi _2 \phi \right) \left(\frac{\phi ^n}{m^n+\phi ^n}\right)^{2 q}-36 n \xi _1 \xi _2 q m^n \text{sech}^2\left(\xi _2 \phi \right) \left(\frac{\phi ^n}{m^n+\phi ^n}\right)^q\nonumber\\&&\left(-2 \xi _2 \phi  \left(m^n+\phi ^n\right) \tanh \left(\xi _2 \phi \right)+m^n (2 n q-1)-(n+1) \phi ^n\right)-96 n \xi _1^3 \xi _2^3 q m^n \text{sech}^6\left(\xi _2 \phi \right) \left(\frac{\phi ^n}{m^n+\phi ^n}\right)^{3 q}\bigg(2 \xi _2 \phi  \left(m^n+\phi ^n\right)\nonumber\\ 
    && \text{tanh} \left(\xi _2 \phi \right)-m^n-(n+1) \phi ^n\bigg)\bigg)\bigg(64 n \xi _1^4 \xi _2^4 q \phi  m^n \left(m^n+\phi ^n\right) \text{sech}^8\left(\xi _2 \phi \right) \left(\frac{\phi ^n}{m^n+\phi ^n}\right)^{4 q}+27 n q \phi  m^n \left(m^n+\phi ^n\right)\nonumber\\&&+192 \xi _1^2 \xi _2^3 \phi ^2 \left(m^n+\phi ^n\right)^2 \tanh \left(\xi _2 \phi \right) \text{sech}^4\left(\xi _2 \phi \right) \left(\frac{\phi ^n}{m^n+\phi ^n}\right)^{2 q}-36 n \xi _1 \xi _2 q m^n \text{sech}^2\left(\xi _2 \phi \right) \left(\frac{\phi ^n}{m^n+\phi ^n}\right)^q\bigg(-2 \xi _2 \phi  \left(m^n+\phi ^n\right)\nonumber\\&& \tanh \left(\xi _2 \phi \right)+m^n (2 n q-1)-(n+1) \phi ^n\bigg)-96 n \xi _1^3 \xi _2^3 q m^n \text{sech}^6\left(\xi _2 \phi \right) \left(\frac{\phi ^n}{m^n+\phi ^n}\right)^{3 q}\bigg(2 \xi _2 \phi  \left(m^n+\phi ^n\right) \tanh \left(\xi _2 \phi \right)-m^n\nonumber\\&&-(n+1) \phi ^n\bigg)\bigg) \bigg)\bigg]\bigg/\bigg[\left(4 \xi _1 \xi _2 \phi  \left(m^n+\phi ^n\right) \text{sech}^2\left(\xi _2 \phi \right) \left(\frac{\phi ^n}{m^n+\phi ^n}\right)^q+3 n q m^n\right)\left(8 \xi _1^2 \xi _2^2 \text{sech}^4\left(\xi _2 \phi \right) \left(\frac{\phi ^n}{m^n+\phi ^n}\right)^{2 q}+3\right)\nonumber\\&&\left(8 \xi _1^2 \xi _2^2 \phi  \left(m^n+\phi ^n\right) \text{sech}^4\left(\xi _2 \phi \right) \left(\frac{\phi ^n}{m^n+\phi ^n}\right)^{2 q}-12 n \xi _1 \xi _2 q m^n \text{sech}^2\left(\xi _2 \phi \right) \left(\frac{\phi ^n}{m^n+\phi ^n}\right)^q+9 \phi  \left(m^n+\phi ^n\right)\right)^2\nonumber\\&&\bigg(27 n q \phi  m^n \left(m^n+\phi ^n\right)+64 n \xi _1^4 \xi _2^4 q \phi  m^n \left(m^n+\phi ^n\right) \text{sech}^8\left(\xi _2 \phi \right) \left(\frac{\phi ^n}{m^n+\phi ^n}\right)^{4 q}+192 \xi _1^2 \xi _2^3 \phi ^2 \left(m^n+\phi ^n\right)^2 \tanh \left(\xi _2 \phi \right) \text{sech}^4\left(\xi _2 \phi \right)\nonumber\\&& \left(\frac{\phi ^n}{m^n+\phi ^n}\right)^{2 q}-36 n \xi _1 \xi _2 q m^n \text{sech}^2\left(\xi _2 \phi \right) \left(\frac{\phi ^n}{m^n+\phi ^n}\right)^q \left(-2 \xi _2 \phi  \left(m^n+\phi ^n\right) \tanh \left(\xi _2 \phi \right)+m^n (2 n q-1)-(n+1) \phi ^n\right)\nonumber\\&&-96 n \xi _1^3 \xi _2^3 q m^n \text{sech}^6\left(\xi _2 \phi \right) \left(\frac{\phi ^n}{m^n+\phi ^n}\right)^{3 q} \left(2 \xi _2 \phi  \left(m^n+\phi ^n\right) \tanh \left(\xi _2 \phi \right)-m^n-(n+1) \phi ^n\right)\bigg)\bigg]
\end{eqnarray}

\subsection{\underline{New Slow Roll Approximation II}}

\label{app2}
The slow-roll parameters in the case of slow roll approximation II are given below:
\begin{eqnarray}
    \delta_1=\frac{2 \xi _1 \xi _2 \text{sech}^4\left(\xi _2 \phi \right) \left(\frac{\phi ^n}{m^n+\phi ^n}\right)^q \left(8 \xi _1 \xi _2 \phi  \left(m^n+\phi ^n\right) \left(\frac{\phi ^n}{m^n+\phi ^n}\right)^q+6 n q m^n \cosh ^2\left(\xi _2 \phi \right)\right)}{18 n \xi _1 \xi _2 q m^n \text{sech}^2\left(\xi _2 \phi \right) \left(\frac{\phi ^n}{m^n+\phi ^n}\right)^q-9 \phi  \left(m^n+\phi ^n\right)}
\end{eqnarray}

\begin{eqnarray}
    \delta_2 &=& \Bigg(-6 n \xi _1 \xi _2 q m^n \text{sech}^2\left(\xi _2 \phi \right) \left(\frac{\phi ^n}{m^n+\phi ^n}\right)^q+16 \xi _1^2 \xi _2^2 \phi  \left(m^n+\phi ^n\right) \text{sech}^4\left(\xi _2 \phi \right) \left(\frac{\phi ^n}{m^n+\phi ^n}\right)^{2 q}+9 \phi  \left(m^n+\phi ^n\right)  \nonumber \\ &&  \Bigg( -3 n q m^n \left(-2 \xi _2 \phi  \left(m^n+\phi ^n\right) \tanh \left(\xi _2 \phi \right)+m^n (n q-1)-(n+1) \phi ^n\right) +8 n q m^n \left(\frac{\phi ^n}{m^n+\phi ^n}\right)^{2 q} \times \nonumber \\&&  \xi _1^2 \xi _2^2 \text{sech}^4\left(\xi _2 \phi \right) \left(-2 \xi _2 \phi  \left(m^n+\phi ^n\right) \tanh \left(\xi _2 \phi \right)+m^n (n q+1)+(n+1) \phi ^n\right)+ 8 \phi  \left(m^n+\phi ^n\right) \text{sech}^2\left(\xi _2 \phi \right) \left(\frac{\phi ^n}{m^n+\phi ^n}\right)^q  \nonumber \\ && \times  \xi _1 \xi _2 \left(2 \xi _2 \phi  \left(m^n+\phi ^n\right) \tanh \left(\xi _2 \phi \right)-n q m^n\right)  \Bigg)\Bigg) \Bigg/ 27 \Bigg(\phi  \left(m^n+\phi ^n\right)-2 n \xi _1 \xi _2 q m^n \text{sech}^2\left(\xi _2 \phi \right) \left(\frac{\phi ^n}{m^n+\phi ^n}\right)^q\Bigg){}^3
\end{eqnarray}

\begin{eqnarray}
 \varepsilon_1 &=& \Bigg( \left( 4 \xi _1 \xi _2 \phi  \left(m^n+\phi ^n\right) \text{sech}^2\left(\xi _2 \phi \right) \left(\frac{\phi ^n}{m^n+\phi ^n}\right)^q+3 n q m^n \right) \Bigg(9 n q \phi ^2 m^n \left(m^n+\phi ^n\right)^2-12 n q \phi  m^n \left(\frac{\phi ^n}{m^n+\phi ^n}\right)^q  \nonumber \\&& \xi _1 \xi _2 \left(m^n+\phi ^n\right) \text{sech}^2\left(\xi _2 \phi \right) \left(-2 \xi _2 \phi  \left(m^n+\phi ^n\right) \tanh \left(\xi _2 \phi \right)+m^n (3 n q-1)-(n+1) \phi ^n\right) -  32 n q \phi  m^n \left(m^n+\phi ^n\right) \nonumber \\ && \left(\frac{\phi ^n}{m^n+\phi ^n}\right)^{3 q} \xi _1^3 \xi _2^3 \text{sech}^6\left(\xi _2 \phi \right) \left(2 \xi _2 \phi  \left(m^n+\phi ^n\right) \tanh \left(\xi _2 \phi \right)-m^n-(n+1) \phi ^n\right) +4 \xi _1^2 \xi _2^2 \text{sech}^4\left(\xi _2 \phi \right) \left(\frac{\phi ^n}{m^n+\phi ^n}\right)^{2 q} \nonumber \\ && n q m^n \left(m^{2 n} \left(3 n^2 q^2-4 \phi ^2\right)-8 m^n \phi ^{n+2}-4 \phi ^{2 n+2}\right)+16 \xi _2 \phi ^3 \left(m^n+\phi ^n\right)^3 \tanh \left(\xi _2 \phi \right) \Bigg)   \Bigg) \Bigg/  \Bigg( 54 \phi  \left(m^n+\phi ^n\right)\nonumber \\ && \left(\phi  \left(m^n+\phi ^n\right)-2 n \xi _1 \xi _2 q m^n \text{sech}^2\left(\xi _2 \phi \right) \left(\frac{\phi ^n}{m^n+\phi ^n}\right)^q\right){}^3 \Bigg)   
\end{eqnarray}

  \begin{align}
     \varepsilon_2 & = -\Bigg[ \Bigg( 3 \text{sech}^2\left(\xi _2 \phi \right) \left(8 \xi _1 \xi _2 \phi  \left(m^n+\phi ^n\right) \left(\frac{\phi ^n}{m^n+\phi ^n}\right)^q+6 n q m^n \cosh ^2\left(\xi _2 \phi \right)\right) \nonumber \\ & \quad \big(6 n \xi _1 \xi _2 q m^n \text{sech}^2\left(\xi _2 \phi \right) \left(\frac{\phi ^n}{m^n+\phi ^n}\right)^q-16 \xi _1^2 \xi _2^2 \phi  \left(m^n+\phi ^n\right) \text{sech}^4\left(\xi _2 \phi \right) \left(\frac{\phi ^n}{m^n+\phi ^n}\right)^{2 q}-9 \phi  \left(m^n+\phi ^n\right) \big) \nonumber \\ & \quad \Bigg( n \phi ^n \left(\phi  \left(m^n+\phi ^n\right)-2 n \xi _1 \xi _2 q m^n \text{sech}^2\left(\xi _2 \phi \right) \left(\frac{\phi ^n}{m^n+\phi ^n}\right)^q\right) \left(4 \xi _1 \xi _2 \phi  \left(m^n+\phi ^n\right) \text{sech}^2\left(\xi _2 \phi \right) \left(\frac{\phi ^n}{m^n+\phi ^n}\right)^q+3 n q m^n\right) \nonumber \\ & \quad \Big( 9 n q \phi ^2 m^n \left(m^n+\phi ^n\right)^2-12 n \xi _1 \xi _2 q \phi  m^n \left(m^n+\phi ^n\right) \text{sech}^2\left(\xi _2 \phi \right) \left(\frac{\phi ^n}{m^n+\phi ^n}\right)^q \nonumber \\ & \quad  \left( -2 \xi _2 \phi  \left(m^n+\phi ^n\right) \tanh \left(\xi _2 \phi \right)+m^n (3 n q-1)-(n+1) \phi ^n \right) -32 n \xi _1^3 \xi _2^3 q \phi  m^n \left(m^n+\phi ^n\right) \text{sech}^6\left(\xi _2 \phi \right) \left(\frac{\phi ^n}{m^n+\phi ^n}\right)^{3 q} \nonumber \\ & \quad \left(  2 \xi _2 \phi  \left(m^n+\phi ^n\right) \tanh \left(\xi _2 \phi \right)-m^n-(n+1) \phi ^n\right) + 4 \xi _1^2 \xi _2^2 \text{sech}^4\left(\xi _2 \phi \right) \left(\frac{\phi ^n}{m^n+\phi ^n}\right)^{2 q} \times \nonumber \\ & \quad \left(n q m^n \left(m^{2 n} \left(3 n^2 q^2-4 \phi ^2\right)-8 m^n \phi ^{n+2}-4 \phi ^{2 n+2}\right)+16 \xi _2 \phi ^3 \left(m^n+\phi ^n\right)^3 \tanh \left(\xi _2 \phi \right)  \right) \Big) + \nonumber \\ &\quad \left(m^n+\phi ^n\right) \left(\phi  \left(m^n+\phi ^n\right)-2 n \xi _1 \xi _2 q m^n \text{sech}^2\left(\xi _2 \phi \right) \left(\frac{\phi ^n}{m^n+\phi ^n}\right)^q\right) \left(4 \xi _1 \xi _2 \phi  \left(m^n+\phi ^n\right) \text{sech}^2\left(\xi _2 \phi \right) \left(\frac{\phi ^n}{m^n+\phi ^n}\right)^q+3 n q m^n\right) \nonumber \\ & \quad \Big( 9 n q \phi ^2 m^n \left(m^n+\phi ^n\right)^2-12 n \xi _1 \xi _2 q \phi  m^n \left(m^n+\phi ^n\right) \text{sech}^2\left(\xi _2 \phi \right) \left(\frac{\phi ^n}{m^n+\phi ^n}\right)^q \nonumber \\& \quad  \left( -2 \xi _2 \phi  \left(m^n+\phi ^n\right) \tanh \left(\xi _2 \phi \right)+m^n (3 n q-1)-(n+1) \phi ^n  \right)- 32 n \xi _1^3 \xi _2^3 q \phi  m^n \left(m^n+\phi ^n\right) \text{sech}^6\left(\xi _2 \phi \right) \left(\frac{\phi ^n}{m^n+\phi ^n}\right)^{3 q} \nonumber \\ & \quad  \left( 2 \xi _2 \phi  \left(m^n+\phi ^n\right) \tanh \left(\xi _2 \phi \right)-m^n-(n+1) \phi ^n \right) + 4 \xi _1^2 \xi _2^2 \text{sech}^4\left(\xi _2 \phi \right) \left(\frac{\phi ^n}{m^n+\phi ^n}\right)^{2 q} \nonumber \\ & \quad \left( n q m^n \left(m^{2 n} \left(3 n^2 q^2-4 \phi ^2\right)-8 m^n \phi ^{n+2}-4 \phi ^{2 n+2}\right)+16 \xi _2 \phi ^3 \left(m^n+\phi ^n\right)^3 \tanh \left(\xi _2 \phi \right) \right)  \Big) + 4 \phi  \left(m^n+\phi ^n\right) \left(\frac{\phi ^n}{m^n+\phi ^n}\right)^q \nonumber \\ & \quad \xi _1 \xi _2 \text{sech}^2\left(\xi _2 \phi \right) \left(\phi  \left(m^n+\phi ^n\right)-2 n \xi _1 \xi _2 q m^n \text{sech}^2\left(\xi _2 \phi \right) \left(\frac{\phi ^n}{m^n+\phi ^n}\right)^q\right) \nonumber\\ & \quad \left( 2 \xi _2 \phi  \left(m^n+\phi ^n\right) \tanh \left(\xi _2 \phi \right)-m^n (n q+1)-\left((n+1) \phi ^n\right) \right) \Big( 9 n q \phi ^2 m^n \left(m^n+\phi ^n\right)^2 - \nonumber \\ & \quad 12 n \xi _1 \xi _2 q \phi  m^n \left(m^n+\phi ^n\right) \text{sech}^2\left(\xi _2 \phi \right) \left(\frac{\phi ^n}{m^n+\phi ^n}\right)^q \left( -2 \xi _2 \phi  \left(m^n+\phi ^n\right) \tanh \left(\xi _2 \phi \right)+m^n (3 n q-1)-(n+1) \phi ^n \right)- \nonumber \\ & \quad 32 n \xi _1^3 \xi _2^3 q \phi  m^n \left(m^n+\phi ^n\right) \text{sech}^6\left(\xi _2 \phi \right) \left(\frac{\phi ^n}{m^n+\phi ^n}\right)^{3 q} \left( 2 \xi _2 \phi  \left(m^n+\phi ^n\right) \tanh \left(\xi _2 \phi \right)-m^n-(n+1) \phi ^n \right) \nonumber \\ & \quad + 4 \xi _1^2 \xi _2^2 \text{sech}^4\left(\xi _2 \phi \right) \left(\frac{\phi ^n}{m^n+\phi ^n}\right)^{2 q} \left(n q m^n \left(m^{2 n} \left(3 n^2 q^2-4 \phi ^2\right)-8 m^n \phi ^{n+2}-4 \phi ^{2 n+2}\right)+16 \xi _2 \phi ^3 \left(m^n+\phi ^n\right)^3 \tanh \left(\xi _2 \phi \right)\right) \Big) \nonumber \\ & \quad + 3 \phi  \left(4 \xi _1 \xi _2 \phi  \left(m^n+\phi ^n\right) \text{sech}^2\left(\xi _2 \phi \right) \left(\frac{\phi ^n}{m^n+\phi ^n}\right)^q+3 n q m^n\right) \times \nonumber \\ & \quad \left(m^n+\phi ^n\right) \left(m^n+(n+1) \phi ^n\right)-\frac{2 n \xi _1 \xi _2 q m^n \text{sech}^2\left(\xi _2 \phi \right) \left(\frac{\phi ^n}{m^n+\phi ^n}\right)^q \left(n q m^n-2 \xi _2 \phi  \left(m^n+\phi ^n\right) \tanh \left(\xi _2 \phi \right)\right)}{\phi } \nonumber \\ & \quad \Big( 9 n q \phi ^2 m^n \left(m^n+\phi ^n\right)^2-12 n \xi _1 \xi _2 q \phi  m^n \left(m^n+\phi ^n\right) \text{sech}^2\left(\xi _2 \phi \right) \left(\frac{\phi ^n}{m^n+\phi ^n}\right)^q \nonumber \\ & \quad \times \left( -2 \xi _2 \phi  \left(m^n+\phi ^n\right) \tanh \left(\xi _2 \phi \right)+m^n (3 n q-1)-(n+1) \phi ^n  \right)- 32 n \xi _1^3 \xi _2^3 q \phi  m^n \left(m^n+\phi ^n\right) \text{sech}^6\left(\xi _2 \phi \right) \left(\frac{\phi ^n}{m^n+\phi ^n}\right)^{3 q} \times \nonumber \\ & \quad \left( 2 \xi _2 \phi  \left(m^n+\phi ^n\right) \tanh \left(\xi _2 \phi \right)-m^n-(n+1) \phi ^n \right)+ 4 \xi _1^2 \xi _2^2 \text{sech}^4\left(\xi _2 \phi \right) \left(\frac{\phi ^n}{m^n+\phi ^n}\right)^{2 q} \times  \nonumber \\ & \quad \left(  \left.n q m^n \left(m^{2 n} \left(3 n^2 q^2-4 \phi ^2\right)-8 m^n \phi ^{n+2}-4 \phi ^{2 n+2}\right)+16 \xi _2 \phi ^3 \left(m^n+\phi ^n\right)^3 \tanh \left(\xi _2 \phi \right)\right) \right)  \Big) - \nonumber \\ & \quad 2 \phi  \left(\phi  \left(m^n+\phi ^n\right)-2 n \xi _1 \xi _2 q m^n \text{sech}^2\left(\xi _2 \phi \right) \left(\frac{\phi ^n}{m^n+\phi ^n}\right)^q\right) \left(4 \xi _1 \xi _2 \phi  \left(m^n+\phi ^n\right) \text{sech}^2\left(\xi _2 \phi \right) \left(\frac{\phi ^n}{m^n+\phi ^n}\right)^q+3 n q m^n\right) \nonumber \\ & \quad  \Bigg( -6 n^2 \xi _1 \xi _2 q^2 m^{2 n} \phi ^n \text{sech}^2\left(\xi _2 \phi \right) \left(\frac{\phi ^n}{m^n+\phi ^n}\right)^{q-1}+9 n^2 q m^n \phi ^{n+1} \left(m^n+\phi ^n\right)^2+9 n q \phi  m^n \left(m^n+\phi ^n\right)^3 \nonumber \\ & \quad \left( -2 \xi _2 \phi  \left(m^n+\phi ^n\right) \tanh \left(\xi _2 \phi \right)+m^n (3 n q-1)-(n+1) \phi ^n \right)  - 6 n^2 \xi _1 \xi _2 q m^n \phi ^{2 n} \text{sech}^2\left(\xi _2 \phi \right) \left(\frac{\phi ^n}{m^n+\phi ^n}\right)^{q-1}\nonumber \\ & \quad \left( -2 \xi _2 \phi  \left(m^n+\phi ^n\right) \tanh \left(\xi _2 \phi \right)+m^n (3 n q-1)-(n+1) \phi ^n \right)-6 n \xi _1 \xi _2 q m^n \left(m^n+\phi ^n\right)^2 \text{sech}^2\left(\xi _2 \phi \right) \left(\frac{\phi ^n}{m^n+\phi ^n}\right)^q \nonumber \\ & \quad \left( -2 \xi _2 \phi  \left(m^n+\phi ^n\right) \tanh \left(\xi _2 \phi \right)+m^n (3 n q-1)-(n+1) \phi ^n  \right)+ 12 n \xi _1 \xi _2^2 q \phi  m^n \left(m^n+\phi ^n\right)^2 \tanh \left(\xi _2 \phi \right) \text{sech}^2\left(\xi _2 \phi \right) \left(\frac{\phi ^n}{m^n+\phi ^n}\right)^q \nonumber \\ & \quad \left( -2 \xi _2 \phi  \left(m^n+\phi ^n\right) \tanh \left(\xi _2 \phi \right)+m^n (3 n q-1)-(n+1) \phi ^n  \right)+ 16 n^2 \xi _1^3 \xi _2^3 q m^n \phi ^{2 n} \text{sech}^6\left(\xi _2 \phi \right) \left(\frac{\phi ^n}{m^n+\phi ^n}\right)^{3 q-1} \nonumber \\ & \quad \left( -2 \xi _2 \phi  \left(m^n+\phi ^n\right) \tanh \left(\xi _2 \phi \right)+m^n+(n+1) \phi ^n \right)+ 16 n \xi _1^3 \xi _2^3 q m^n \left(m^n+\phi ^n\right)^2 \text{sech}^6\left(\xi _2 \phi \right) \left(\frac{\phi ^n}{m^n+\phi ^n}\right)^{3 q} \nonumber \\ & \quad \left( -2 \xi _2 \phi  \left(m^n+\phi ^n\right) \tanh \left(\xi _2 \phi \right)+m^n+(n+1) \phi ^n\right)- 96 n \xi _1^3 \xi _2^4 q \phi  m^n \left(m^n+\phi ^n\right)^2 \tanh \left(\xi _2 \phi \right) \text{sech}^6\left(\xi _2 \phi \right) \left(\frac{\phi ^n}{m^n+\phi ^n}\right)^{3 q} \nonumber \\ & \quad \left(-2 \xi _2 \phi  \left(m^n+\phi ^n\right) \tanh \left(\xi _2 \phi \right)+m^n+(n+1) \phi ^n \right) - 48 n^2 \xi _1^3 \xi _2^3 q^2 m^{2 n} \phi ^n \text{sech}^6\left(\xi _2 \phi \right) \left(\frac{\phi ^n}{m^n+\phi ^n}\right)^{3 q-1}  \nonumber \\ & \quad \left( 2 \xi _2 \phi  \left(m^n+\phi ^n\right) \tanh \left(\xi _2 \phi \right)-m^n-(n+1) \phi ^n  \right) + 16 \xi _1^2 \xi _2^2 \phi  \left(m^n+\phi ^n\right)^2 \text{sech}^4\left(\xi _2 \phi \right) \left(\frac{\phi ^n}{m^n+\phi ^n}\right)^{2 q} \nonumber \\ & \quad \big( 6 \xi _2 \phi  \left(m^n+\phi ^n\right)^2 \tanh \left(\xi _2 \phi \right)+6 n \xi _2 \phi ^{n+1} \left(m^n+\phi ^n\right) \tanh \left(\xi _2 \phi \right)+2 \xi _2^2 \phi ^2 \left(m^n+\phi ^n\right)^2 \text{sech}^2\left(\xi _2 \phi \right)-n q m^n \left(m^n+(n+1) \phi ^n\right)  \big) \nonumber \\ & \quad + \frac{4 n \xi _1^2 \xi _2^2 q m^n \text{sech}^4\left(\xi _2 \phi \right) \left(\frac{\phi ^n}{m^n+\phi ^n}\right)^{2 q} \left(n q m^n \left(m^{2 n} \left(3 n^2 q^2-4 \phi ^2\right)-8 m^n \phi ^{n+2}-4 \phi ^{2 n+2}\right)+16 \xi _2 \phi ^3 \left(m^n+\phi ^n\right)^3 \tanh \left(\xi _2 \phi \right)\right)}{\phi } \nonumber \\ & \quad - 8 \xi _1^2 \xi _2^3 \phi ^n \tanh \left(\xi _2 \phi \right) \text{sech}^4\left(\xi _2 \phi \right) \left(\frac{\phi ^n}{m^n+\phi ^n}\right)^{2 q-1} \times
  \nonumber \\ & \quad \big( n q m^n \left(m^{2 n} \left(3 n^2 q^2-4 \phi ^2\right)-8 m^n \phi ^{n+2}-4 \phi ^{2 n+2}\right)+16 \xi _2 \phi ^3 \left(m^n+\phi ^n\right)^3 \tanh \left(\xi _2 \phi \right) \big) + \nonumber \\ & \quad 16 n \xi _1^3 \xi _2^3 q \phi  m^n \left(m^n+\phi ^n\right)^2 \text{sech}^6\left(\xi _2 \phi \right) \left(\frac{\phi ^n}{m^n+\phi ^n}\right)^{3 q} \times \nonumber \\ & \quad \big( -2 \xi _2 \left(m^n+(n+1) \phi ^n\right) \tanh \left(\xi _2 \phi \right)-2 \xi _2^2 \phi  \left(m^n+\phi ^n\right) \text{sech}^2\left(\xi _2 \phi \right)+n (n+1) \phi ^{n-1} \big)+ 6 n q \phi  m^n \left(m^n+\phi ^n\right)^2 \left(\frac{\phi ^n}{m^n+\phi ^n}\right)^q \times \nonumber \\ & \quad \xi _1 \xi _2 \text{sech}^2\left(\xi _2 \phi \right) \big( 2 \xi _2 \left(m^n+(n+1) \phi ^n\right) \tanh \left(\xi _2 \phi \right)+2 \xi _2^2 \phi  \left(m^n+\phi ^n\right) \text{sech}^2\left(\xi _2 \phi \right)+n (n+1) \phi ^{n-1} \big) \Bigg) \Bigg)  \Bigg) \Bigg/ \nonumber \\ & \quad  \Big( 2 \phi  \left(m^n+\phi ^n\right) \left(9 \phi  \left(m^n+\phi ^n\right)-18 n \xi _1 \xi _2 q m^n \text{sech}^2\left(\xi _2 \phi \right) \left(\frac{\phi ^n}{m^n+\phi ^n}\right)^q\right){}^2 \times \nonumber \\ & \quad \big( \phi  \left(m^n+\phi ^n\right)-2 n \xi _1 \xi _2 q m^n \text{sech}^2\left(\xi _2 \phi \right) \left(\frac{\phi ^n}{m^n+\phi ^n}\right)^q \big) \big( 4 \xi _1 \xi _2 \phi  \left(m^n+\phi ^n\right) \text{sech}^2\left(\xi _2 \phi \right) \left(\frac{\phi ^n}{m^n+\phi ^n}\right)^q+3 n q m^n\big) \nonumber \\ & \quad \big( 9 n q \phi ^2 m^n \left(m^n+\phi ^n\right)^2-12 n \xi _1 \xi _2 q \phi  m^n \left(m^n+\phi ^n\right) \text{sech}^2\left(\xi _2 \phi \right) \left(\frac{\phi ^n}{m^n+\phi ^n}\right)^q \times \nonumber \\ & \quad \left( -2 \xi _2 \phi  \left(m^n+\phi ^n\right) \tanh \left(\xi _2 \phi \right)+m^n (3 n q-1)-(n+1) \phi ^n \right)- 32 n \xi _1^3 \xi _2^3 q \phi  m^n \left(m^n+\phi ^n\right) \text{sech}^6\left(\xi _2 \phi \right) \left(\frac{\phi ^n}{m^n+\phi ^n}\right)^{3 q} \times \nonumber \\ & \quad  \left( 2 \xi _2 \phi  \left(m^n+\phi ^n\right) \tanh \left(\xi _2 \phi \right)-m^n-(n+1) \phi ^n \right) + 4 \xi _1^2 \xi _2^2 \text{sech}^4\left(\xi _2 \phi \right) \left(\frac{\phi ^n}{m^n+\phi ^n}\right)^{2 q} \times \nonumber \\ & \quad \big(  n q m^n \left(m^{2 n} \left(3 n^2 q^2-4 \phi ^2\right)-8 m^n \phi ^{n+2}-4 \phi ^{2 n+2}\right)+16 \xi _2 \phi ^3 \left(m^n+\phi ^n\right)^3 \tanh \left(\xi _2 \phi \right)\big) \big)  \Big)     \Bigg] 
  \end{align}

\subsection{Field values in Slow Roll approximations and Numerical analysis}

\begin{table}[h]
\centering
\begingroup
\setlength{\tabcolsep}{10pt} 
\renewcommand{\arraystretch}{1.5} 

\begin{minipage}{0.45\textwidth}
\centering
\caption{Table for field value at start of inflation ($\phi_{\rm i}$) and end of inflation ($\phi_{\rm e}$) for different values of potential parameters when  $\Delta N = 60$ in Slow-Roll I}
\label{phi_tabslow1}
\begin{tabular}{ |c|c|c|c| } 
\hline
$q$ & $n$ & $\phi_{\rm i}$ & $\phi_{\rm e}$ \\
\hline
    & 1 & 8.8622 & 0.8987 \\
 1   & 2 & 8.6941 & 1.2516 \\
    & 3 & 8.6167 & 1.3598 \\
     & 4 & 8.5966 & 1.3836 \\
\hline
\hline
    & 1 & 9.1927 & 1.4516 \\
 2   & 2 & 8.8128 & 1.7501 \\
    & 3 & 8.6536 & 1.7205 \\
     & 4 & 8.6131 & 1.6510 \\
\hline
\hline
     & 1 & 9.5287 & 1.7497 \\
 3   & 2 & 8.9218 & 2.0874 \\
    & 3 & 8.6833 & 1.9572 \\
     & 4 & 8.6243 & 1.8212 \\
\hline
\hline
      & 1 & 9.8713 & 1.8422 \\
 4   & 2 & 9.0266 & 2.3325 \\
    & 3 & 8.7102 & 2.1378 \\
     & 4 & 8.6333 & 1.9498 \\
\hline
  
\end{tabular}
\end{minipage}%
\hspace{0.05\textwidth}
\begin{minipage}{0.45\textwidth}
\centering
\caption{Table for field value at start of inflation ($\phi_{\rm i}$) and end of inflation ($\phi_{\rm e}$) for different values of potential parameters when  $\Delta N = 60$ in Slow-Roll II}
\label{phi_tabslow2}

\begin{tabular}{ |c|c|c|c| } 
\hline
$q$ & $n$ & $\phi_{\rm i}$ & $\phi_{\rm e}$ \\
\hline
    & 1 & 8.9850 & 3.3626 \\
 1   & 2 & 8.7967 & 3.5634 \\
    & 3 & 8.7181 & 3.5727 \\
     & 4 & 8.7009 & 3.5579 \\
\hline
\hline
    & 1 & 9.2961 & 3.1765 \\
 2   & 2 & 8.8963 & 3.5800 \\
    & 3 & 8.7391 & 3.6010 \\
     & 4 & 8.7048 & 3.5744 \\
\hline
\hline
     & 1 & 9.6231 & 2.9812 \\
 3   & 2 & 8.9955 & 3.5919 \\
    & 3 & 8.7601 & 3.6261 \\
     & 4 & 8.7088 & 3.5897 \\
\hline
\hline
      & 1 & 9.9600 & 2.7781 \\
 4   & 2 & 9.0942 & 3.6001 \\
    & 3 & 8.7309 & 3.6487 \\
     & 4 & 8.7127 & 3.6041 \\
\hline
\end{tabular}
\end{minipage}

\endgroup
\end{table}

\begin{table}[h!] 
\centering
\caption{Table for field value at start of inflation ($\phi_{\rm i}$) and end of inflation ($\phi_{\rm e}$) for different values of potential parameters when  $\Delta N = 60$  obtained by numerically solving Eq.~\ref{DynSYSN}.}
\label{phi_tab_numeric}
\renewcommand{\arraystretch}{1.5} 
\setlength{\tabcolsep}{12pt} 
\begin{tabular}{ |c|c|c|c| } 
\hline
$q$ & $n$ & $\phi_{\rm i}$ & $\phi_{\rm e}$ \\
\hline
    & 1 & 8.9022 & 0.4532 \\
 1   & 2 & 8.6732 & 0.8638 \\
    & 3 & 8.6013 & 1.0831 \\
     & 4 & 8.5851 & 1.1783 \\
\hline
\hline
    & 1 & 9.1311 & 0.6928 \\
 2   & 2 & 8.7753 & 1.1894 \\
    & 3 & 8.6273 & 1.3588 \\
     & 4 & 8.5934 & 1.4015 \\
\hline
\hline
     & 1 & 9.4492 & 0.8829 \\
 3   & 2 & 8.8875 & 1.3964 \\
    & 3 & 8.6513 & 1.5272 \\
     & 4 & 8.6000 & 1.5341 \\
\hline
\hline
      & 1 & 9.7838 & 1.0458 \\
 4   & 2 & 8.9744 & 1.5573 \\
    & 3 & 8.6743 & 1.6533 \\
     & 4 & 8.6060 & 1.6304 \\
\hline
\end{tabular}
\end{table}

\end{document}